\documentclass[fleqn,usenatbib]{mnras}

\usepackage{newtxtext,newtxmath}
\usepackage[T1]{fontenc}

\DeclareRobustCommand{\VAN}[3]{#2}
\let\VANthebibliography\thebibliography
\def\thebibliography{\DeclareRobustCommand{\VAN}[3]{##3}\VANthebibliography}

\usepackage[normalem]{ulem}

\usepackage{graphicx}	
\usepackage{amsmath}	

\newcommand{\mcO}{\mathcal{O}}
\newcommand{\mcM}{\mathcal{M}}

\newcommand{\kmsperMpc}{{\rm km}\,{\rm s}^{-1}\,{\rm Mpc}^{-1}}

\newcommand{\gcmc}{{\rm g\; cm}^{-3}}

\newcommand{\Zsolar}{\;{\rm Z}_{\odot}}
\newcommand{\msolar}{{\rm M}_{\odot}}

\newcommand{\OVI}{\hbox{O\,{\sc vi}}}
\newcommand{\OVII}{\hbox{O\,{\sc vii}}}
\newcommand{\OVIII}{\hbox{O\,{\sc viii}}}

\newcommand{\HI}{{\hbox{H\,{\sc i}}}}

\newcommand{\nh}{{n_{\rm H}}}

\title[The Descriptive Parametric Model]{Introducing the Descriptive Parametric Model: Gaseous Profiles for Galaxies, Groups, and Clusters}

\author[B. D. Oppenheimer et al.]{
\parbox[t]{\textwidth}{
Benjamin D. Oppenheimer,$^{1}$\thanks{E-mail: benjamin.oppenheimer@colorado.edu}
G. Mark Voit,$^{2}$
Yannick M.~Bah\'e,$^{3,4}$ 
Nicolas Battaglia,$^{5,6}$
Joel Bregman,$^{7}$
Joseph N. Burchett,$^{8}$
Dominique Eckert,$^{9}$
Yakov Faerman,$^{10}$
Justus Gibson,$^{1}$
Cameron Hummels,$^{11}$
Isabel Medlock,$^{12}$
Daisuke Nagai,$^{13,12}$
Mary Putman,$^{14}$
Zhijie Qu,$^{15,16}$
Ming Sun,$^{17}$
Jessica K. Werk,$^{10}$
Yi Zhang$^{18}$}
\\
\\
$^{1}$University of Colorado, Center for Astrophysics and Space Astronomy, 389 UCB, Boulder, CO 80309, USA\\
$^{2}$Michigan State University, Department of Physics and Astronomy, East Lansing, MI 48824, USA\\
$^{3}$School of Physics and Astronomy, University of Nottingham, University Park, Nottingham NG7 2RD, UK\\
$^{4}$Laboratory of Astrophysics, Ecole Polytechnique F\'{e}d\'{e}rale de Lausanne (EPFL), Observatoire de Sauverny, 1290 Versoix, Switzerland\\
$^{5}$Department of Astronomy, Cornell University, Ithaca, NY 14853, USA \\
$^{6}$Université Paris Cité, CNRS, Astroparticule et Cosmologie, F-75013 Paris, France\\
$^{7}$Department of Astronomy University of Michigan Ann Arbor, MI  48109\\
$^{8}$Department of Astronomy, New Mexico State University, Las Cruces, NM 88001, USA\\
$^{9}$Department of Astronomy, University of Geneva, Ch. d'Ecogia 16, CH-1290 Versoix, Switzerland\\
$^{10}$University of Washington, Department of Astronomy, Seattle, WA, 98195\\
$^{11}$TAPIR, California Institute of Technology, Pasadena, CA 91125, USA\\
$^{12}$Department of Astronomy, Yale University, New Haven, CT 06520, USA\\
$^{13}$Department of Physics, Yale University, New Haven, CT 06520, USA\\
$^{14}$Department of Astronomy, Columbia University, New York, NY 10027, USA\\
$^{15}$Department of Astronomy \& Astrophysics, The University of Chicago, 5640 S. Ellis Ave., Chicago, IL 60637, USA\\
$^{16}$Department of Astronomy, Tsinghua University, Beĳing 100084, People’s Republic of China\\
$^{17}$Department of Physics and Astronomy, University of Alabama in Huntsville, 301 Sparkman Dr NW, Huntsville, AL 35899, USA\\
$^{18}$Max-Planck-Institut f\"ur extraterrestrische Physik (MPE), Gie{\ss}enbachstra{\ss}e 1, D-85748 Garching bei M\"unchen, Germany
}

\date{Accepted XXX. Received YYY; in original form ZZZ}

\pubyear{\the\year{}}

\begin{document}
\label{firstpage}
\pagerange{\pageref{firstpage}--\pageref{lastpage}}
\maketitle

\begin{abstract}
We develop and present the Descriptive Parametric Model (DPM), a tool for generating profiles of gaseous halos (pressure, electron density, and metallicity) as functions of radius, halo mass, and redshift. The model assumes single-phase, spherically symmetric, volume-filling warm/hot gas. The DPM framework enables mock observations of the circumgalactic medium (CGM), group halos, and clusters across a number of wavebands including X-ray, sub-millimeter/millimeter, radio, and ultraviolet (UV). We introduce three model families calibrated to reproduce cluster profiles while having different extrapolations to the CGM -- (i) self-similar halos, (ii) a reduced gas model for lower halo masses, and (iii) a model with shallower radial slopes at lower masses. We demonstrate how our $z=0.0-0.6$ models perform when applied to stacked and individual X-ray emission profiles, measurements of the thermal and kinetic Sunyaev-Zel'dovich Effect, electron dispersion measures from fast radio bursts, $\OVI$ absorption, and UV-derived pressures. Our investigation supports models that remove baryons from halos more effectively and have shallower profiles at lower halo mass. We discuss biases and systematics when modelling observables using consistent hot gaseous halo models for all wavebands explored. We release the {\tt DPMhalo} code to encourage the use of our framework and new formulations in future investigations.  Included with the {\tt DPMhalo} distribution is a set of recent observations that allow the reproduction of most plots in this paper.
\end{abstract}


\section{Introduction}\label{sec:intro}

Despite remarkable progress on multiple observational fronts, a unified description of the fundamental physical properties---pressure, density, and metallicity---of the circumgalactic medium (CGM) remains elusive. Even interpretations of the Milky Way's CGM, the one we know the most about, suffer from limitations, in part because we often do not know the distance to the circumumgalactic gas we are detecting \citep{sembach2003,Shull2009,Lehner2011,putman2012,Zheng2019,Locatelli2024}. Furthermore, the volume-filling component of the CGM around other galaxies similar to the Milky Way has yet to be characterized out to radii similar to the halo's virial radius \citep[e.g.][]{Bogdan2013,Li2017}.  

At the very least, we would like to express the physical properties of the CGM as one-dimensional functions of radius, out to the virial radius and perhaps even beyond it, and we expect those properties to depend on halo mass and to evolve with redshift. From this perspective, the CGM is a volume-filling, spherical {\it atmosphere}, similar to the gaseous halos of galaxy clusters and groups,\footnote{In this paper, we call halos with masses $M_{200} \geq 10^{14}\;\msolar$ galaxy clusters and halos with masses $10^{13}-10^{14}\;\msolar$ galaxy groups.} which both X-ray observations \citep[e.g.][]{Pratt2009,Lovisari2021} and observations of the Sunyaev-Zel'dovich (SZ) Effect \citep[e.g.][]{Planck2014,Bleem2015} show to be volume-filling and often nearly spherical. Modelling the CGM as a one-dimensional, spherically symmetric atmosphere is admittedly simplistic, but it is useful as a starting point that can be tested through comparisons with observations.   

While data revealing the atmospheres of clusters and groups are abundant, detections of diffuse atmospheres around typical galaxies in lower mass halos ($\la 10^{13}\;\msolar$) are still uncertain. Ultraviolet (UV) absorption line data provide most of what we know about these gaseous halos, but might not be telling us very much about the volume-filling phase.  The neutral hydrogen ($\HI$) and low-ionization metals detected in UV spectra are best modelled as a cool ($\sim 10^4-10^5\;$K) component \citep[e.g.][]{Stocke2013,werk2014,Chen2018}, often inferred to arise from {\it cloud}-like structures occupying a small fraction of the CGM's volume.  Indeed, many hydrodynamic simulations of galaxies like the Milky Way produce photo-ionized CGM clouds embedded within a hotter volume-filling atmosphere \citep[e.g.][]{joung12, Gutcke2017,opp18c,Hummels2019,Peeples2019,Vandevoort2019,Nelson2020,Ramesh2023}.  

In this paper, we aim to connect observations of cloud properties inferred from UV observations with X-ray and SZ observations of higher-mass halos using a tool that we call the Descriptive Parametric Model (DPM). It assumes that galactic atmospheres over the entire range of halo masses we explore are spherical, single-phase, and have a relatively simple dependence on radius.  One main goal is to provide new constraints on the CGM around typical galaxies, in halos belonging to the mass range $M_{200}\approx 10^{11.5}-10^{13}\;\msolar$, which we will call ``$L^\star$ halos." 

Importantly, the DPM does not assume a particular physical scenario for the CGM. It is \textit{descriptive by design}, in contrast with models that make physical assumptions like precipitation \citep{Voit2017,Voit2019_PrecipLimited}, isothermal profiles \citep{Faerman2017}, interior virial shocks \citep{Stern2018}, cooling gas \citep{Qu2018a,Qu2018b}, isentropic profiles \citep{Faerman2020,Faerman2022}, rotating coranae \citep{Pezzulli2017,Sormani2018}, or rotating cooling flows \citep{Stern2024}.  It is inspired by observational datasets and follows just a few theoretical guidelines to make it physically plausible for galaxy halos.  It is therefore meant for \textit{empirical fitting} and can be applied independently of theoretical considerations.  

The purposes of this first DPM paper are therefore four-fold:
\begin{enumerate}
    \item To introduce the mathematical formalism of the DPM (\S\ref{sec:formalism}).  
    \item To develop the tools necessary to derive mock observational data from the DPM (\S\ref{sec:mock}).
    \item To compile the initial version of a gaseous halo data library for making comparisons with the DPM over a large range in halo mass.
    \item To demonstrate how the DPM reveals patterns of similarity and tension among the existing observational data sets. 
\end{enumerate}
We have made the DPM an open-access community tool for exploring how CGM properties depend on the halo mass and redshift.  

Section \ref{sec:formalism} fulfills the first purpose by introducing the formalism describing three key properties of gaseous halos: gas pressure $P$, electron density $n_\mathrm{e}$, and gas metallicity $Z$.  The DPM consists of radial profiles for each property that depend on physical radius $r$, halo mass $M$, and redshift $z$.  Those radial profiles are not simple power laws for reasons explained in \S\ref{sec:formalism}. 

Section \ref{sec:mock} addresses the second purpose, explaining how mock observables in the following wavebands are derived from the DPM:
\begin{itemize}
    \item \textbf{X-rays} (\S\ref{sec:mockXray}) --- Soft X-ray emission is detected from individual groups and clusters, but galactic halos often require stacking techniques.  
    \item \textbf{Sub-mm/mm} (\S\ref{sec:mockSZ}) --- SZ observations measure electron density and pressure, usually through stacking of $L^\star$ and group halos.
    \item \textbf{Radio} (\S\ref{sec:mockFRB}) --- Dispersion measures obtained via fast radio bursts (FRBs) provide a direct probe of the intervening electron density.  
    \item \textbf{UV} (\S\ref{sec:mockUV}) --- Absorption line spectroscopy offers multiple means to measure gaseous atmospheres. The $\OVI$ doublet traces $\sim3\times 10^5$ K gas in its collisionally ionized phase (\S\ref{sec:mockOVI}).  Low-ionization metal lines provide the opportunity to fit physical models that return a measurement of pressure (\S\ref{sec:mockPressures}).\footnote{UV observations are discussed last owing to this waveband's complex systematics, combined with large and diverse datasets.}  
\end{itemize}
We discuss how we simulate each of the above observables and include descriptions of the associated software packages.  

Purposes three and four are addressed in Sections \ref{sec:models} through \ref{sec:implications}, where we demonstrate how the DPM can be applied.  In \S\ref{sec:models}, we introduce three different implementations of the DPM that all reproduce the well-observed radial profiles of galaxy clusters, but have different behaviours at lower mass. We begin first with cluster profiles simply scaled down to $L^\star$ halos while keeping the halo baryon fraction constant (\S\ref{sec:model1}).  Second, we explore profiles that keep the same radial dependence but have baryon fractions that decline as halo mass decreases (\S\ref{sec:model2}).  The third and final model has profiles that vary both in shape and baryon fraction with halo mass (\S\ref{sec:model3}). We discuss the physical properties that we directly model (\S\ref{sec:primephys}) and the secondary physical properties that are indirectly modelled (\S\ref{sec:secondphys}). We recognize that the first model (for certain) and the second model (quite probably) are already known {\it not} to fit all of the available data.  
We are presenting them here to demonstrate how the DPM can converge toward a more unified description by revealing what needs to be adjusted.

Section \ref{sec:obs} compares DPM profiles to the observations compiled in our data repository.  We discuss what each observable measures, which involves determining whether an observable measures the atmospheric component that we are modelling as a 1-D profile, the cloud-like and non-spherical structures embedded within the atmosphere, or a combination of both.  We begin with the X-ray (\S\ref{sec:obsXray}) as these datasets provide the most guidance for our initial DPMs, followed by SZ (\S\ref{sec:obsSZ}), FRBs (\S\ref{sec:obsDM}), and finally determine how the DPMs compare with UV datasets (\S\ref{sec:obsUV}).  Section \ref{sec:implications} discusses the implications for physical profiles from multi-waveband constraints using DPM profiles.  Section \ref{sec:applications} addresses additional applications of the {\tt DPMhalo} module, and Section \ref{sec:summary} summarizes the paper. 

This paper is written such that sections are individually accessible.  For example, the formalism (\S\ref{sec:formalism}) and the generation of mock observables (\S\ref{sec:mock}) are written for a reader interested in applying the DPM to their research.  Alternatively, a reader may access the description of our initial models (\S\ref{sec:model1}-\ref{sec:model3}), their physical properties (\S\ref{sec:primephys}-\ref{sec:secondphys}), and how they compare to observations, including a discussion of each observable's potential biases (\S\ref{sec:obs}) followed by potential tensions across multiple wavebands (\S\ref{sec:implications}).  

The default cosmological model for the DPM is a \citet[][]{Planck2018}  cosmology with parameters $\Omega_M=0.311$, $\Omega_b=0.0490$, $H_0=67.66 \;\kmsperMpc$, referred to as the Planck 2018 cosmology.  Subscripts indicating multiples of overdensity, e.g.~$R_{200}$, $M_{500}$, refer to the critical rather than mean density; we indicate the latter as $R_{200,m}$.  When converting between different radii and masses, we use the {\tt Colossus} package \citep{diemer2018} with the mass-concentration relationship of \citet{diemer2015}. 


\section{Descriptive Parametric Model Formalism} \label{sec:formalism}

The Descriptive Parametric Model (DPM) consists of a set of radial profiles that depend on physical radius $r$, halo mass $M$, and redshift $z$.  We model the radial profiles of three properties: gas pressure $P$, electron density $n_\mathrm{e}$, and gas metallicity $Z$.  
Previous CGM studies often parametrized the gas thermodynamic profiles with some simplifying assumptions and a minimal set of parameters (e.g. \citealp{Miller2013,Faerman2017,Voit2017,Stern2019,Oren2024}).  However, we opted to emphasize flexibility over simplicity, and therefore added more parameters to account for curvature of the radial profiles.  The rationale behind this is that better future data will allow more complex radial profiles to be fit to the CGM of galaxies.

Two other factors motivated the decision for a more complex parametrization.  First, pressure and density profiles of observed clusters and groups have curved profiles with different power-law slopes at inner, transition, and outer radii \citep[e.g.][]{Arnaud2010,McDonald2017}.  Second, modern cosmological simulations of clusters and groups show radial profiles with curvature \citep[e.g.][]{opp21,braspenning2024}.  We therefore use the generalized NFW (gNFW) profile of \citet{Nagai2007}, who simultaneously fitted simulated and observed cluster profiles.  


All of our profiles depend on radius according to the generalized NFW profile function
\begin{equation}
    f(x|\boldsymbol{\alpha}) \equiv x^{-\alpha_{\rm in}} (1 + x^{\alpha_{\rm tr}})^{(\alpha_{\rm in}-\alpha_{\rm out})/\alpha_{\rm tr}}
    \; \; ,
\end{equation}
where $x \equiv r / R_{\rm s}$ is the radius $r$ in units of a scale radius $R_{\rm s}$, and $\boldsymbol{\alpha} = (\alpha_{\rm in}, \alpha_{\rm tr}, \alpha_{\rm out})$ is a vector of power-law parameters.\footnote{This is the \citet{Nagai2007} gNFW form with $\gamma$, $\alpha$, and $\beta$
replaced by subscripted $\alpha$ coefficients with abbreviations for inner, transition, and outer radii.} At radii much smaller than $R_{\rm s}$, the radial profile function converges to $f(x) \propto x^{-\alpha_{\rm in}}$. At radii much larger than $R_{\rm s}$, it converges to $f(x) \propto x^{-\alpha_{\rm out}}$. The power law slope changes near $R_{\rm s}$, and the parameter $\alpha_{\rm tr}$ controls how rapidly the slope changes.  To fix the value of $R_{\rm s}$, we set $c_{200}$, defined as $R_{200}/R_{\rm s}$, to a single value of $2.772$ for all the profiles.  We find that this parameter is effectively degenerate with combinations of the $\alpha$ parameters for the gNFW profiles.  We choose this value because it is the concentration favored by the \citet{Nagai2007} best-fit to the pressure profiles of simulated galaxy clusters: $c_{500}=1.8$ and $c_{200}/c_{500}=1.54$.

Using this function of radius, the DPM pressure profile becomes
\begin{equation}
     P(r,M,z) = P_0 \, f(x|\boldsymbol{\alpha}^P) E(z)^{\gamma^P} M_{12}^{\beta^P} 
     \; \; ,\label{equ:P}
\end{equation}
where $P_0$ is a pressure-normalization parameter, $M_{12} \equiv M_{200} / 10^{12} \, M_\odot$ represents the halo mass, $E(z) = [\Omega_{\rm M}(1+z)^3 + \Omega_\Lambda]^{1/2}$ is the unitless Hubble parameter, $\boldsymbol{\alpha}^P = (\alpha_{\rm in}^P, \alpha_{\rm out}^P, \alpha_{\rm tr}^P)$ is the vector of radial power-law parameters for pressure, and the parameters $\beta_P$ and $\gamma_P$ represent the power-law dependences on $M_{200}$ and $E(z)$, respectively. We note that this formula is a slightly modified version of the \citet{Nagai2007} pressure profile, in that it is scaled to $R_{200}$ instead of $R_{500}$, the radius inside which the mean density is $500 \times$ the critical density.   

Likewise, the DPM electron density and metallicity profiles are given by
\begin{equation}
     n_\mathrm{e}(r,M,z) = n_{\mathrm{e}0} \, f(x|\boldsymbol{\alpha}^{n_\mathrm{e}}) E(z)^{\gamma^{n_\mathrm{e}}} M_{12}^{\beta^{n_\mathrm{e}}} \label{equ:ne}
\end{equation}
and
\begin{equation}
     Z(r,M,z) = Z_0 \, f(x|\boldsymbol{\alpha}^Z) E(z)^{\gamma^Z} M_{12}^{\beta^Z}
     \; \; . \label{equ:Z}
\end{equation}

Noting that the radial profile shapes differ as a function of mass, we introduce mass dependence to the radial profile exponents, 
\begin{equation}
\alpha^P_{\rm out} = \alpha^P_{\rm out,12}+\alpha^P_{\rm out,var} \times {\rm log}_{10}(M/M_{12}),\label{equ:Mdep_radial}
\end{equation}
where $\alpha^P_{{\rm out},12}$ is the outer radial pressure exponent at $M=10^{12}\;\msolar$ and $\alpha^P_{\rm out,var}$ sets the mass dependence of the outer radial pressure exponent.  This is repeated for every radial exponent for $P$, $n_\mathrm{e}$, and $Z$.  

We also add a scatter parameter $\sigma^{n_\mathrm{e}}$, which we define in dex, to account for log-normal isobaric fluctuations in electron density around the mean $\bar{n}_e$ at a given radius.  The probability distribution of $n_\mathrm{e}$ 
is given by 
\begin{equation}
    p(n_\mathrm{e}|\mu,\sigma) 
        =     \frac{1}{n_\mathrm{e} \sigma \sqrt{2 \pi}} {\rm exp}\left[ -\frac{({\rm ln}\;n_\mathrm{e}-\mu)^2}{2 \sigma^2} \right], 
        \label{equ:lognormal_PDF}
\end{equation}
which has a log-normal dispersion $\sigma = \sigma^{n_\mathrm{e}} \times {\rm ln}\;10$ and a mean $\mu$ of ${\rm ln}\;n_\mathrm{e}$ that equals $\mu = {\rm ln}\; \bar{n}_e - \sigma^2/2$.  The integral of Equ. \ref{equ:lognormal_PDF} is the cumulative distribution function of $n_\mathrm{e}$,
\begin{equation}
    \frac{1}{2} \left[ 1 + {\rm erf} \left( \frac{{\rm ln}\;n_\mathrm{e} - \mu}{\sqrt{2}\sigma}  \right)  \right].
\end{equation}
Temperatures, with a mean $\bar{T}$, have corresponding inverted fluctuations to preserve isobaric conditions.
A log-normal scatter has also been used by \citet{Dutta2024} to model the multi-phase Milky Way CGM.  

We do not add scatter parameters for metallicity due to these measurements being less constrained via fewer observational datasets.  Neither do we add scatter for pressure for reasons that include i) our approach that emphasizes simpler pressure profiles that allow tests of physical principles including hydrostatic equilibrium, and ii) that such fluctuations would manifest as fluctuations in temperature given our formalism, which could lead to confusion with $\sigma^{n_\mathrm{e}}$ that has a reflective scatter in $T$.  However, additional scatter parameters could be added in a future implementation. 

The parameter $r_{\rm max}$ sets the 3-dimensional (3-D) radius out to which the profiles are generated.  We set this parameter to $3 R_{200}$, but allow a user other choices in generating their profiles out to different radii. 

Altogether, a particular implementation of the DPM is specified by the set of parameters  used in Equations \ref{equ:P}$-$\ref{equ:Z}: $\Theta \equiv \{ P_{0}, \alpha^P_\mathrm{out,12}, \alpha^P_\mathrm{out,var}, ...., \gamma_{Z}, r_{\rm max} \}$.  There are 32 parameters in all, including $\sigma^{n_\mathrm{e}}$ and $r_{\rm max}$. For each physical property, seven parameters are required to specify the radial dependence, and three more give the property's normalization, redshift dependence, and halo-mass dependence.  In practice, our code uses normalization inputs of the profile at $0.3 R_{200}$ ($P_{0.3}$, $n_{\mathrm{e},0.3}$, $Z_{0.3}$). In this paper, each set $\mcM$ of model profiles comprises $P(r,M,z)$, $\bar{n}_e(r,M,z)$, and $Z(r,M,z)$, and each profile assumes $r_{\rm max} = 3 R_{200}$.

\section{Mocking Observational Datasets}\label{sec:mock}

We use the DPM formalism to create mock observable profiles projected onto the sky.  To generate a particular mock observable $\bar{\mcO}_i$, we calculate the appropriate volume-weighted emission or absorption density $\Upsilon_i(r | \Theta )$, which is a function of radius depending on a particular DPM model set $\mcM (\Theta)$ and therefore on a particular DPM parameter set $\Theta$.  We translate between 3-D radial ($r$) and projected mock observed profiles ($r_{\perp}$) by integrating localized volumes along intersecting rays through the halo using an Abel projection: 
\begin{equation}
    \bar{\mcO}_i(r_{\perp}| \Theta) = 2 \int^{r_{\rm max}}_{r_{\perp}}  
    \Upsilon_i(r | \Theta) \, \frac{r \,  dr}{\sqrt{r^2-r_{\perp}^2}}.  \label{equ:Abel} 
\end{equation}
Each profile is generated out to a projected radius $r_{\perp} = 2 R_{200}$ using all gas within $r_{\rm max}$.  

We introduce observational datasets divided into four wavebands that probe gaseous halos: X-ray (\S\ref{sec:mockXray}), submillimeter/millimeter (\S\ref{sec:mockSZ}), radio (\S\ref{sec:mockFRB}), and UV (\S\ref{sec:mockUV}).  We briefly comment on how each observable relates to the atmospheric component that is the target of our modelling.  We then explain how we generate each mock observation. 

\subsection{Soft X-ray Emission} \label{sec:mockXray}

X-ray emission can map gaseous atmospheres, but is rarely detected beyond the galactic disk around typical galaxies \citep[e.g.][]{anderson2016,bogdan2017,Li2017,das2020}.   Stacking X-ray halos around many galaxies is now possible from the {\it eROSITA} mission \citep{Predehl2021}, even using a small patch of the sky in the early release eFEDS survey \citep{comparat2022,chadayammuri2022}.  More recently \citet{Zhang2024a,Zhang2024b} performed {\it eROSITA} stacking using the 2-year eRASS:4 Western Galactic Hemisphere dataset, mapping emission extending out to the virial radius.  We can use these stacks as loose constraints, noting that non-atmospheric X-ray emission likely contributes to these stacks since deep imaging of individual galactic halos finds non-spherical structures.  In particular, X-ray emission arises from what appear to be wind-blown bipolar cavities around other galaxies \citep[e.g.][]{HodgesKluck2020} as well as the {\it eROSITA} bubbles observed above and below our own Milky Way disk \citep{predehl2020}.  We keep this aspect in mind when comparing our models to the \citet{Zhang2024a,Zhang2024b} datasets, which could contain and possibly be dominated by such structures.

X-ray emission is observed around individual clusters and groups, allowing the determination of physical profiles out to large radii \citep[e.g.][]{Finoguenov2002,Ponman2003,morandi2017}.  ICM pressure, density, and metallicity profiles are measured around a uniformly selected sample of nearby clusters from the X-COP program \citep{Eckert2017,Ghirardini2019,Ghizzardi2021}.  Groups have also had their profiles mapped out to large radii in a number of surveys \citep{sun2009,sun2011,lovisari2015,lovisari2019}.  Recognizing the unique value of mapping profiles around individual objects combined with the perspective that the scalings of the ICM and IGrM may well apply to the CGM of typical galaxies, we use cluster and group profiles as essential linchpins in the parametric models presented in this manuscript. 


\subsubsection{Mocking X-ray Observables}

For X-ray emission, we calculate 
\begin{equation}
\Upsilon_\mathrm{Xray}(r) 
= \int^\infty_0 \epsilon_{\rm Xray}[P(r),n_\mathrm{e},Z(r)] \, p[n_\mathrm{e}|\mu(r),\sigma^{n_\mathrm{e}}] \, dn_\mathrm{e}
\end{equation}
by integrating $\epsilon_{\rm Xray}$, the 0.5-2.0 keV (soft) X-ray volume emissivity given in erg s$^{-1}$ cm$^{-3}$, across the probability distribution of densities (and inversely temperatures) according to the log-normal distribution in Equ. \ref{equ:lognormal_PDF}.  

We use the {\tt pyXSIM} code \citep{Zuhone2016} that accesses lookup tables for $\epsilon_{\rm Xray}$ calculated using CLOUDY \citep{Ferland2017} by \citet{khabibullin2019}, which depend on $n_\mathrm{e}$, $T$, $Z$, and $z$.  
We integrate $\Upsilon_\mathrm{Xray}$ via Equ. \ref{equ:Abel} to calculate $0.5-2.0$ keV surface brightness in units of erg s$^{-1}$ kpc$^{-2}$.  Although {\tt pyXSIM} has the capability to generate photon lists via a Monte Carlo method, we turn this feature off for the initial set of models presented here.

\subsection{Sunyaev-Zel'dovich Effects} \label{sec:mockSZ}

The submillimeter/millimeter waveband allows the measurement of gas properties via the thermal and kinetic Sunyaev-Zel'dovich Effects (abbreviated as tSZ and kSZ) that can provide determinations of electron pressure and density, respectively.  This is a mapping technique using Compton scattering of the Cosmic Microwave Background (CMB) radiation, and requires complex stacking methods to detect gaseous halos for all but the most massive clusters \citep{Bleem2015,Hilton2021}. \citet{Schaan2021} stacked over $3\times 10^5$ halos corresponding to galaxy groups at $z\sim 0.4-0.7$ yielding tSZ-derived pressure and kSZ-derived density profiles, which were modelled by a companion paper \citep{Amodeo2021}.  At $z\sim 0$ and using $\sim 10$ objects, \citet{pratt2021} and \citet{bregman2022} stack galaxy groups and spiral galaxies, respectively, finding tSZ detections in both cases.  

The SZ Effect provides direct measurements of the physical properties of thermal pressure and free electron density that do not depend on metallicity.  All ionized gas is included in these measurements, but only group and cluster mass objects, where the atmospheric component dominates, have kSZ measurements.  The pressure, which we model uniformly across cloud-like structures and atmospheres, is measured via tSZ. 

\subsubsection{Mocking the SZ Effect}

For tSZ, we calculate 
\begin{equation}
\Upsilon_\mathrm{tSZ} = P(r) \times \frac{k \sigma_{\rm T}}{m_e c^2} 
\end{equation}
at each radial position, where $k$ is the Boltzmann constant, $\sigma_{\rm T}$ the Thomson cross-section, $m_e$ electron mass, and $c$ the speed of light.  We project $\Upsilon_\mathrm{tSZ}$ along the line of sight via Equ. \ref{equ:Abel} to return the projected mock observable $\bar{\mcO}_{tSZ}$, which is the unitless Compton $y$ parameter at a given $r_{\perp}$ and is the quantity that \citet{pratt2021} and \citet{bregman2022} plot.  The tSZ Effect is also quantified via CMB temperature fluctuations given in units of micro-Kelvin, and measured within an angular distance $\theta$, as described in eqs. 3 and 4 of \citet{Amodeo2021}, including an expression for the frequency dependence.  Hence, these measurements, $T_{tSZ}$, are temperature decrements multiplied by enclosed angular area and typically have units of $\mu$K arcmin$^2$, which is a cumulative measurement.  

For kSZ, the corresponding calculation is
\begin{equation}
\Upsilon_\mathrm{kSZ} = \bar{n}_e(r) \times \sigma_{\rm T},
\end{equation}
which becomes the Thomson optical depth, $\tau$, upon projection via Equ. \ref{equ:Abel}.  This is converted into a temperature fluctuation within an angular distance as described by Equ. 5 and 6 of \citet{Amodeo2021}.  

We use the {\tt Mop-c-GT} package\footnote{https://github.com/samodeo/Mop-c-GT} from \citet{Amodeo2021} to mock the ACT (Atacama Cosmology Telescope) measurements cross-correlated with the BOSS CMASS halos of \citet{Schaan2021}.  This package includes the above calculations for tSZ and kSZ, treatment of stacked halos of various masses, as well as the convolution with the ACT beam.  Within {\tt Mop-c-GT}, we use DPM profiles out to $3\times R_{200}$ to represent the 1-halo term and add in the \citet{Amodeo2021} 2-halo term (see their fig. 7 for a representation).    

\subsection{Electron Dispersion Measures} \label{sec:mockFRB}

A nascent measure of ionized electrons comes from radio telescopes that are now capable of detecting thousands of FRBs at cosmological distances, providing a dispersion measure (DM) due to free electrons along sight lines toward the emitted FRB.  A localized FRB (i.e. an FRB with a known host galaxy and redshift) provides a measurement of the entire electron column density between our position and the emitting object.  These DMs can measure most of the Universe's baryon content under the safe assumption that most of cosmic baryons are ionized \citep{MacQuart2020}.  Early attempts to cross-correlate nearby halos with FRBs provide DM estimates as a function of the CGM impact parameter \citep{Wu2023}. DMs can powerfully constrain the ionized baryon content of gaseous halos once the contamination from electrons in and around our Galaxy as well as the distant FRB host galaxy are corrected for \citep{medlock2024a}.  

DM quantities include ionized gas in the atmosphere as well as cloud structures as long as they are ionized.  Lower mass halos may well have a significant fraction of ionized cloud-like structures; therefore, the DM values can be treated as an upper limit for the atmospheric component.  

\subsubsection{Calculating Dispersion Measures}

The dispersion measure ``density'' is given by
\begin{equation}
\Upsilon_\mathrm{DM} = \frac{\bar{n}_e(r)}{1+z}
\end{equation}
and is integrated using Equ. \ref{equ:Abel}, resulting in dispersion measures having units of pc$\;$cm$^{-3}$.  \citet{Wu2023} provides DM as a function of radius around stacked galaxies, and we do not attempt to reproduce their data reduction procedure.  




\subsection{Ultraviolet Absorption} \label{sec:mockUV}

The ultraviolet (UV) waveband currently provides the most knowledge about the CGM in $L^*$ halos.  UV absorption is the best existing method to measure diffuse gas properties at typical CGM radii \citep{tumlinson2017}.  Multiple significant observing programs primarily using the Cosmic Origins Spectrograph (COS) on the {\it Hubble Space Telescope} have measured numerous ionic species in the ``low'' redshift ($z\la$0.6) Universe.  We note in particular the surveys of COS-Halos \citep{werk2014,Prochaska2017}, COS-LRG \citep{Chen2018,Zahedy2019}, and COS-GTO \citep{Stocke2013,Keeney2017}, as these datasets contain two separate observable constraints: $\OVI$ absorption strengths and UV-derived pressures from embedded clouds. 

\subsubsection{$\OVI$ Absorption} \label{sec:mockOVI}

The first UV observational constraint is from the $\OVI$ ion observed in absorption along quasar sight lines.  Around star-forming galaxies, the $\OVI$ column density appears very uniform relative to $\HI$ and low ions, and then drops off considerably for more passive galaxies \citep{tumlinson2011}.  The more complete CGM$^2$ survey \citep{Tchernyshyov2022} combining this and other surveys \citep[e.g.][]{Johnson2015,Keeney2017,keeney2018,Zahedy2019} indicates an $\OVI$ profile around typical galaxies that is strong and steeply declines approaching $R_{200}$.
It is therefore worthwhile to consider $\OVI$ as a possible tracer of a volume-filling medium.  

\paragraph{Mocking $\OVI$ Absorption}

We determine $\OVI$ column density, $N_{\OVI}$, by integrating $\Upsilon_{\OVI}(r| \Theta)$ along the line of sight using 

\begin{equation}
\Upsilon_\mathrm{\OVI}(r) 
= \int^\infty_0 n_{\OVI}[P(r),n_\mathrm{e},Z(r)] \, p[n_\mathrm{e}|\mu(r),\sigma^{n_\mathrm{e}}] \, dn_\mathrm{e}
\end{equation}
where $n_{\OVI}$, the number density of $\OVI$ per cm$^{3}$, is calculated from CLOUDY-generated tables for $n_\mathrm{e}$, $T$, and $z$ that account for ionization from the extra-galactic ionization background (EGB for short).  Like X-rays, $n_{\OVI}$ includes a $\sigma^{n_\mathrm{e}}$ dependence.  We use the lookup tables via {\tt Trident} \citep{Hummels2017} using the \citet[][hereafter HM12]{Haardt2012} EGB.  These tables can be scaled linearly by metallicity.  While these tables contain both collisionally ionized (CI) and photo-ionized $\OVI$, our atmosphere models contain mainly the former due to DPM temperatures in our model set being above $10^5$ K in all cases.  {\tt Trident} also provides the \citet[][hereafter FG09]{fauchergiguere2009} EGB that has fewer hard ionizing photons and allows more CI $\OVI$ at lower densities, which we also explore.  

\subsubsection{UV Absorption-Derived Pressures} \label{sec:mockPressures}

The second UV observable is the determination of CGM pressures.  This is a unique probe as we use measures of physical properties of embedded clouds to infer the pressures of those clouds and to assess whether they are in pressure equilibrium with the surrounding atmosphere.  UV absorption line datasets including neutral hydrogen ($\HI$) and aligned low ion metals can be modelled using CLOUDY as $T \sim 10^4$ K gas.  Electron density is determined from CLOUDY models, from which pressure can be determined using $P = n_\mathrm{e}\times T$.  Notably, clouds are treated as having distinct positions as opposed to continuous volume-filling distributions like other observables; hence we cannot easily de-project the 3-D radial position from the observed position of the derived pressure.  Some absorption line systems have multiple components and multiple pressure measurements \citep[e.g.][]{Zahedy2019}, which presumably arise from intersecting multiple clouds at different radii along the sightline.  In such cases, we take the highest pressure assuming it arises at the lowest $r$.  Because pressures are quantified as projected $r_{\perp}$ and we show these observationally-derived quantities on a physical plot as a function of $r$, we indicate with rightward arrows that the clouds may be at a radius $r$ significantly greater than $r_\perp$ (see Figure \ref{fig:pressure_all}).

\paragraph{Calculating Observationally-Derived Pressures}

We begin with the \citet{Voit2019} compilation from various samples including \citet[][COS-Halos]{Prochaska2017},  \citet[][COS-GTO]{Keeney2017,keeney2018}, and \citet[][COS-LRG]{Zahedy2019}. \citet{Voit2019} collects the ionization parameter $U\equiv n_\gamma/n_\mathrm{H}$ where $n_\gamma$ is the number of ionizing photons for neutral hydrogen.  For our assumption of a uniform EGB and a fully ionized medium, $U$ is inversely proportional to $n_\mathrm{e}$.  We slightly update the pressures compiled by \citet{Voit2019} as follows.  The relationship between $U$ and $n_\mathrm{e}$ depends on the EGB used, and the HM12 background is the default EGB.  We additionally explore the older HM05 EGB that is based on the \citet{Haardt2001} EGB but slightly modified when used in CLOUDY, and is several times stronger than the HM12 background at $z\la 0.5$ \citep{Gibson2022}.  There are several more recent calculations of the EGB \citep[e.g.][]{khaire2019, puchwein2019,Fauchergiguere2020}, and the HM05 and HM12 EGBs bracket the range of these other EGBs. We therefore explore these two EGBs, as did \citet{Voit2019}, to consider the systematics within this measurement.  One minor modification from \citet{Voit2019} is that we more precisely calculate the redshift-dependence of the ionizing photon density for both HM12 and HM05.  Another minor modification is the calculation of the temperature required to determine pressure.  Like \citet{Voit2019}, we use the temperature set by the photo-ionization equilibrium with the EGB, but we use the $\nh-T$ relation from CLOUDY equilibrium tables calculated by \citet{opp13a} assuming $Z=0.3\Zsolar$ and the $z=0.2$ HM12 EGB.   These changes are relatively minor with $\sim 0.1$ dex uncertainty for the equilibrium temperature due to metallicity \citep{Faerman2025} and $\sim 0.1$ dex from different EGBs and redshifts \citep{opp13a}.

\subsubsection{Stellar Mass to Halo Mass Conversion}\label{sec:SMHM}

Absorption line measurements are typically parametrized by their projected impact parameter from a galaxy.  The galaxy information most often known are stellar mass and galaxy type.  We use the \citet{Behroozi2019} {\tt UniverseMachine} software to generate stellar mass-to-halo mass (SMHM) conversions.  We have modified a python script using the \citet{Behroozi2019} parameters, fit, and code to flexibly generate SMHM lookup tables for 11 redshifts between $z=0-1$.  We use this software's capability to treat star-forming and quiescent galaxies separately by creating SMHM lookup tables for each category.  The differences between galaxy type is small ($<10\%$, cf. fig. 11 \citet{Behroozi2019}) for the galaxies with absorption measurements.

We generate errors for each SMHM.  The {\tt UniverseMachine} returns $1\;\sigma$ ranges for the SMHM.  We also treat the uncertainty of the stellar mass to be $\pm 0.2$ dex, which propagates through our calculation by adding the $1\;\sigma$ range to our chosen stellar mass uncertainty.

When the galaxy masses are different between the \citet{Voit2019} compilation and the $\OVI$ database of \citet{Tchernyshyov2022}, we choose the lower $M_*$ for both.  We also put an upper limit of $M_\star=10^{11.0}\;\msolar$ on galaxies not in \citet{Tchernyshyov2022}, which corresponds to a halo mass of $\sim 10^{13}\;\msolar$.  We argue that lower stellar masses are more probable due to Eddington bias, i.e. owing to the existence of more lower mass than higher mass halos.  

\section{Initial Use Cases} \label{sec:models}

We now present three initial models using the DPM formalism.  The first two models (\S\ref{sec:model1}-\ref{sec:model2}) are based on cluster profiles that are scaled down to the $L^\star$ regime with different assumptions.  The third (\S\ref{sec:model3}) is a model that allows for the mass-dependent flattening of profiles at lower masses.  These models are ``initial'' DPMs, because more data will become available in the future.  
We demonstrate the pressure, density, and metallicity profiles in \S\ref{sec:primephys} and expand to additional physical properties in \S\ref{sec:secondphys}.

\subsection{Model 1: Self-Similar Scaling} \label{sec:model1}

Model 1 is the base assumption that all halos are self-similar scalings of observed clusters.  We use the results of X-COP, the {\it XMM} Cluster Outskirts Project observing 12 $z<0.1$ massive clusters, for the pressure and density profiles \citep[][hereafter G19]{Ghirardini2019}.  The G19 density profile accounts for almost the entire cosmic abundance of baryons, $f_{\rm b}\equiv \Omega_{\rm b}/\Omega_{\rm M}$, which is expected for massive clusters that are unable to significantly eject baryons via feedback due to their deep potential wells.  This model is known to be unrealistic for $L^\star$ halos since bright X-ray halos predicted around typical galaxies \citep{white1991} were not observed in {\it ROSAT} data \citep{benson2000}.  Nonetheless, we use the G19 pressure and density profiles as the high-mass anchor point for all our models at $M_{200} \sim 10^{15}\;\msolar$.  For metallicity, we assume a uniform slope calibrated to groups from \citet{lovisari2019}.  Model 1 is referred to as the cluster-based ``Self-Similar'' model.  

\subsection{Model 2: Scaling of Fractional Gas Mass} \label{sec:model2}

It is well established that lower mass groups do not contain baryons at the cosmic average $f_{\rm b}$ inside $R_{500}$ \citep{Eckert2021}, therefore Model 2 scales down the gas fraction to calibrate to group gas fractions observed by \citet{Akino2022}, while using the same profile shapes as clusters based on G19.  Pressures are also scaled downward using the \citet{Arnaud2010} mass dependence.  We name this the ``Cluster-Reduced'' model.  Presuming that feedback ejects baryons and rearranges their profiles in lower mass halos, it may be considered physically improbable to expect profiles to maintain the same shape as clusters.  

\subsection{Model 3: Scaling of Profile Slopes} \label{sec:model3}

Our third model alters the profile shape as a function of mass based on several physical criteria.  Model 3, the ``Slope-Changing'' model, is the motivation for the mass-dependent radial formulation in Equ. \ref{equ:Mdep_radial}.  X-ray observations support flatter $P$ and $n_\mathrm{e}$ profiles than clusters in the group regime \citep{sun2009, sun2011, lovisari2015}, but we do not use these results for calibration owing to their samples being X-ray-selected (see \S\ref{sec:obsXraygroups} for further discussion).  We calibrate to the group gas fractions of \citet{Akino2022} at $M_{500}=10^{13.0-13.5}\;\msolar$ using G19 clusters as a basis.  This model's primary assumption is that pressure and density profiles become flatter from the cluster regime through the group regime and into the galactic halo regime.  We apply three physical criteria to achieve flatter profiles:
\begin{itemize}
    \item Gas fractions must be less than $f_{\rm b}$.  
    \item Entropy profiles should not be declining as a function of $r$ inside $R_{500}$ at $M_{200}\ga 10^{12}\;\msolar$.
    \item Cooling times should be comparable or longer than a Hubble time, except at $\ll R_{200}$.  
\end{itemize} 

The first criterion is easily satisfied given the declining trend of density toward lower halo mass.  We additionally enforce this criterion to allow for the mass budgeting of the stellar component as well as the non-spherical/cloud-like component of denser structures embedded within the gaseous atmosphere.  This criterion also results from the expected behaviour of feedback being capable of removing baryons while there is no standard process that adds excess baryons to a halo.  The second criterion provides a challenge as group entropy profiles are shallower than their cluster counterparts; hence extrapolating this to the galaxy scale can create a declining entropy profile that is dynamically unstable.  Model 3 ends up with a nearly flat entropy profile at $10^{12}\;\msolar$, which is the basis of the \citet{Faerman2020} isentropic model that is conceivable if AGN and stellar feedback drive a convective equilibrium throughout the halo.  Finally, the median cooling time must be comparable to or in excess of the Hubble time at radii outside the extent of the central galaxy (see Figure \ref{fig:tcool}).  This requirement is based on hot halos being dynamically stable for a length comparable to the age of the Universe, which we show in \S\ref{sec:secondphys} when discussing cooling times.  We note that this is {\it our} requirement for this model, and may well not be applicable in reality.   

\begin{table}
  \caption{Model Parameters$^{a}$}
  \begin{center}
  \begin{tabular}{lccc}
    \hline
     & Model 1 & Model 2 & Model 3 \\
     & \scriptsize{Self-Similar} & \scriptsize{Cluster-Reduced} & \scriptsize{Slope-Changing} \\

    $P_{0.3}$ & 409 & 115 & 71 \\
    $\alpha^{P}_{\rm in,12}$ & 0.3 & 0.3 & -0.6 \\
    $\alpha^{P}_{\rm tr,12}$ & 1.3 & 1.3 & 0.2 \\
    $\alpha^{P}_{\rm out,12}$ & 4.1 & 4.1 & 2.0 \\
    $\alpha^{P}_{\rm in,var}$ & 0 & 0 & 0.3 \\
    $\alpha^{P}_{\rm tr,var}$ & 0 & 0 & 0.37 \\
    $\alpha^{P}_{\rm out,var}$ & 0 & 0 & 0.7 \\
    $c^{P}_{200}$ & 2.772 & 2.772 & 2.772 \\
    $\beta^{P}$ & 2/3 & 0.85 & 0.92 \\
    $\gamma^{P}$ & 8/3 & 8/3 & 8/3 \\
    $n_{\mathrm{e}_{0.3}}$ & $5.86\times 10^{-4}$ & $4.87\times 10^{-5}$ & $4.87\times 10^{-5}$ \\
    $\alpha^{n_\mathrm{e},12}_{\rm in}$ & 1.0 & 1.0 & 0.4 \\
    $\alpha^{n_\mathrm{e},12}_{\rm tr}$ & 1.9 & 1.9 & 0.45 \\
    $\alpha^{n_\mathrm{e},12}_{\rm out}$ & 2.7 & 2.7 & 0.5 \\
    $\alpha^{n_\mathrm{e}}_{\rm in,var}$ & 0 & 0 & 0.2 \\
    $\alpha^{n_\mathrm{e}}_{\rm tr,var}$ & 0 & 0 & 0.48 \\
    $\alpha^{n_\mathrm{e}}_{\rm out,var}$ & 0 & 0 & 0.73 \\
    $c^{n_\mathrm{e}}_{200}$ & 2.772 & 2.772 & 2.772 \\    
    $\beta^{n_\mathrm{e}}$ & 0 & 0.36 & 0.36 \\
    $\gamma^{n_\mathrm{e}}$ & 2 & 2 & 2 \\
    $\sigma^{n_\mathrm{e}}$ & 0.15 & 0.15 & 0.01,0.15,0.30$^{b}$ \\
    $Z_{0.3}$ & 0.3 & 0.3 & 0.3 \\
    $\alpha^{Z}_{\rm in,12}$ & 0.0 & 0.0 & 0.0 \\
    $\alpha^{Z}_{\rm tr,12}$ & 0.5 & 0.5 & 0.5 \\
    $\alpha^{Z}_{\rm out,12}$ & 0.7 & 0.7 & 0.7 \\
    $\alpha^{Z}_{\rm in,var}$ & 0$^{c}$ & 0$^{c}$ & 0$^{c}$ \\
    $\alpha^{Z}_{\rm tr,var}$ & 0$^{c}$ & 0$^{c}$ & 0$^{c}$ \\
    $\alpha^{Z}_{\rm out,var}$ & 0$^{c}$ & 0$^{c}$ & 0$^{c}$ \\ 
    $c^{Z}_{200}$ & 2.772 & 2.772 & 2.772 \\    
    $\beta^{Z}$ & 0$^{c}$ & 0$^{c}$ & 0$^{c}$ \\
    $\gamma^{Z}$ & 0$^{c}$ & 0$^{c}$ & 0$^{c}$ \\
    $r_{\rm max}$ & $3 R_{200}$ & $3 R_{200}$ & $3 R_{200}$ \\
    \hline    
  \end{tabular}
  \end{center}
  {$^{a}$ Our formulation of the DPM uses these 32 parameters.  The radial slope parameters are normalized at $M=10^{12}\;\msolar$ as noted by the subscript $_{12}$.  Model 3 has mass-dependent slopes for $P$ and $n_\mathrm{e}$ as indicated by non-zero $_\mathrm{var}$ subscripts, while the $_{12}$ subscripts apply to all masses for Models 1 and 2.}\\
  {$^b$ The $\sigma^{n_e}$ log-normal scatter parameter is varied for different instances of Model 3 only.}\\
  {$^c$ The metallicity parameters for varying radial slope as a function of mass ($\alpha^{Z}_{\rm in,var}$, $\alpha^{Z}_{\rm tr,var}$, $\alpha^{Z}_{\rm out,var}$), mass dependence ($\beta^{Z}$), and redshift evolution ($\gamma^{Z}$) are always zero for the Models 1-3, but are included for future applications.}\\
\label{tab:Params}
\end{table}

Table \ref{tab:Params} lists the entire 32 parameter sets $\Theta$ for the three models that are contained in the publicly released {\tt DPMhalo} code.  The radial profile slopes are a combination of $\alpha_{12}$ and $\alpha_\mathrm{var}$ variables according to Equ. \ref{equ:Mdep_radial}.  A singular metallicity profile, independent of mass and redshift, sets the $\alpha^{Z}_{\rm in,var}$, $\alpha^{Z}_{\rm tr,var}$, $\alpha^{Z}_{\rm out,var}$,  $\beta^Z$, and $\gamma^Z$ parameters to zero. 

Lastly, the negative slope for $\alpha^P_{\rm in}$ for Model 3 indicates that this profile asymptotes to a positive pressure gradient in the interior for $L^\star$ halos; however we show in \S\ref{sec:Pphys} that the pressure gradient is always negative at CGM radii.  This also demonstrates that the multiple power laws of the gNFW radial profile do not simply translate into actual power laws for the radii of our focus.  We use the gNFW profile to recreate the smoothly curving profiles seen in simulated and observed profiles.  

\begin{figure*}
\includegraphics[width=0.49\textwidth]{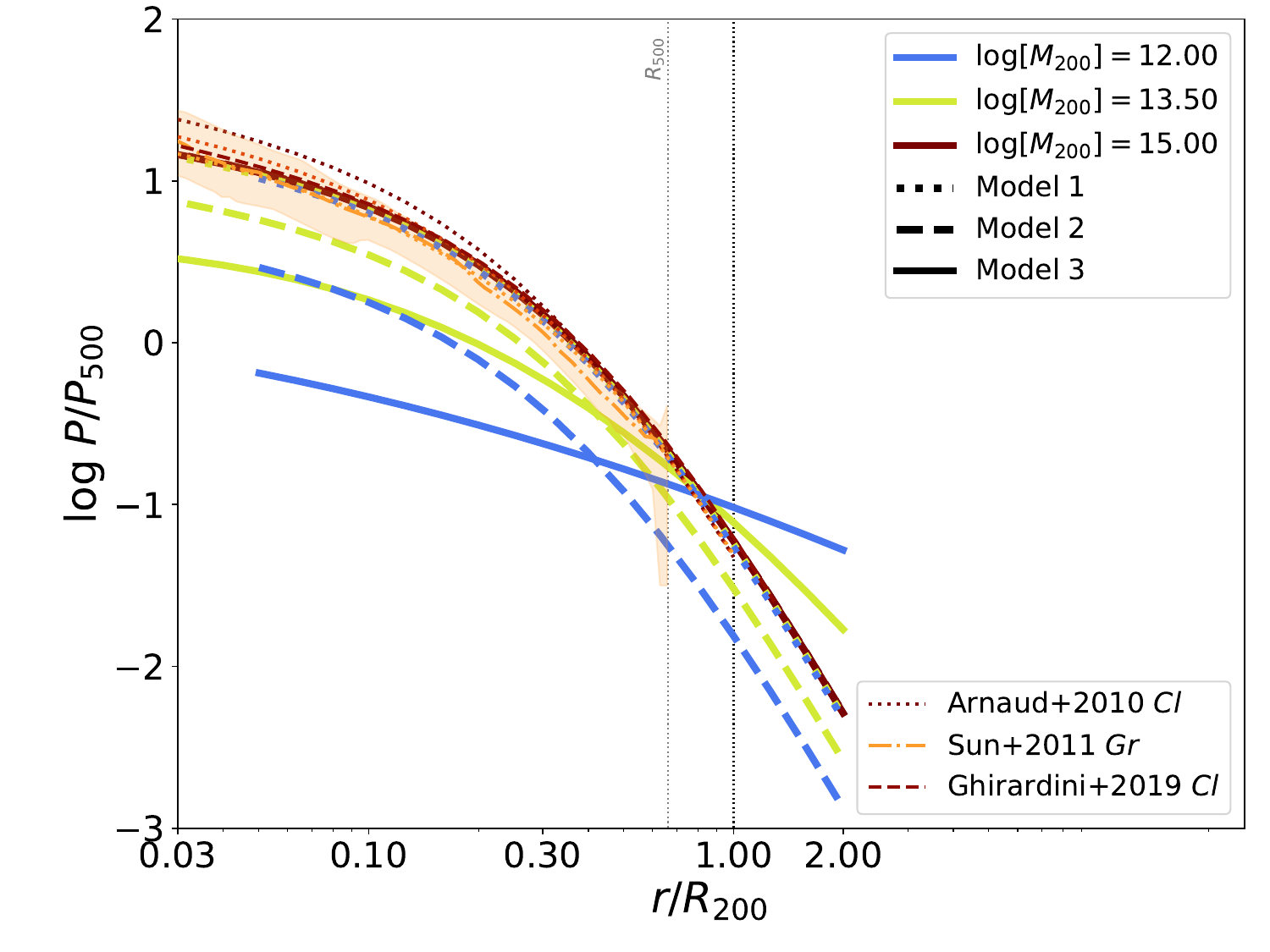}
\includegraphics[width=0.49\textwidth]{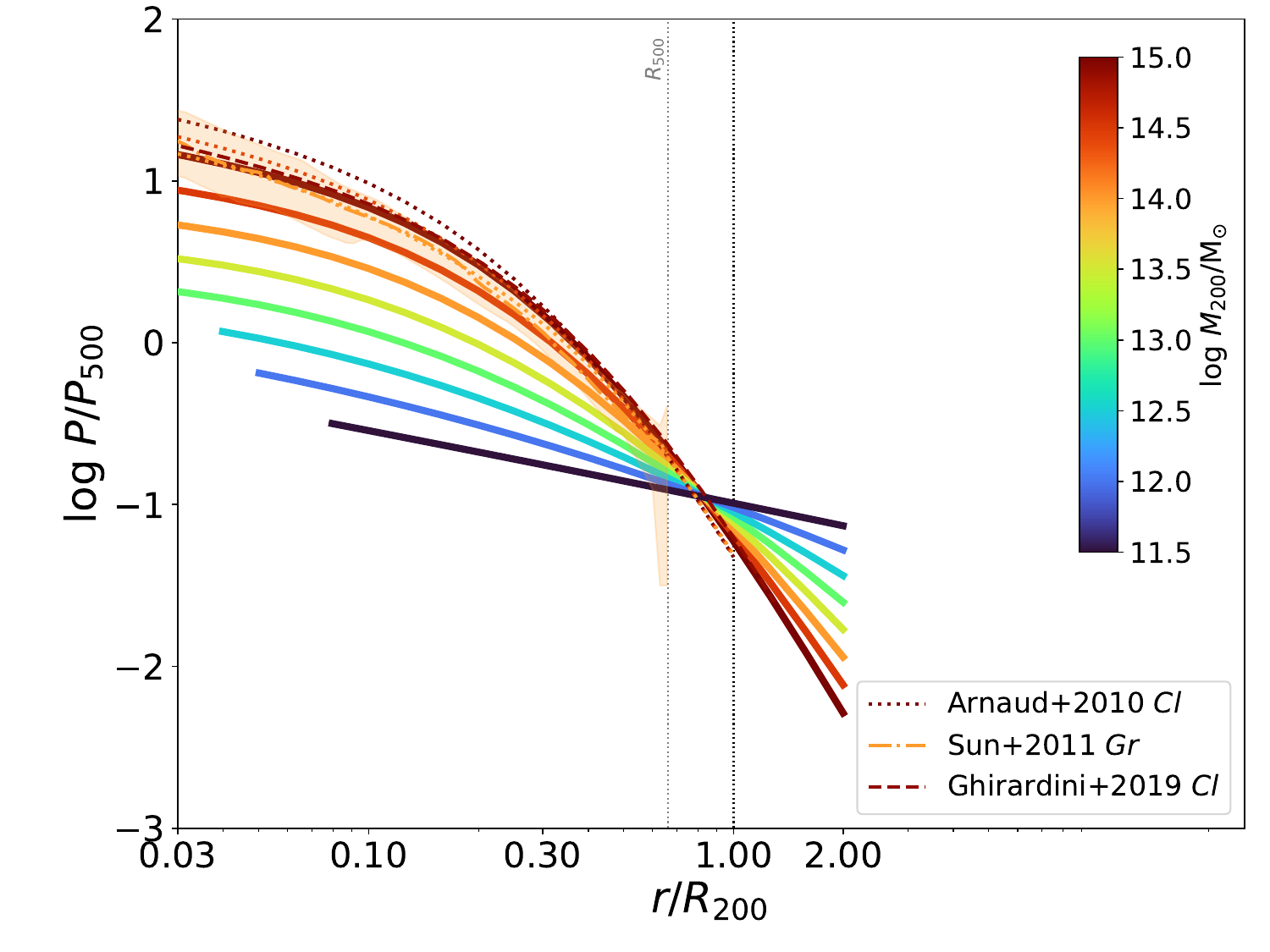}
\caption{{\it Left:} Normalized pressure for the three models at three masses ($L^\star$- blue, Groups- light green, and Clusters- dark red).  Clusters for all models and all profiles from Model 1 overlap the dark red DPM profile. {\it Right:} Mass dependence for Model 3.  The color scheme for halo masses, indicated by the colorbar, is used throughout all figures for DPMs and observational datasets.  The masses shown on the right are log($M_{200}/\msolar$)=11.5, 12.0, 12.5, 13.0, 13.5, 14.0, 14.5, \& 15.0.  The \citet{Arnaud2010} pressure relationships derived from clusters are shown for 3 masses.  The \citet{sun2011} median profile for massive groups and the G19 fitted profile from X-COP clusters are also indicated (the latter is mostly overlapped by the $10^{15.0}\;\msolar$ DPM profiles).  These datasets are more clearly shown in the unnormalized pressure plot in Fig. \ref{fig:pressure_all}. Dotted vertical lines indicating $R_{200}$ and $R_{500}$ are repeated for all physical property figures.}  
\label{fig:pressure_norm}
\end{figure*}

\subsection{Primary Physical Properties} \label{sec:primephys}

\subsubsection{Pressure}\label{sec:Pphys}


The pressure profiles are normalized by $P_{500}\equiv T_{500} n_{\mathrm{e},500}$, where $k T_{500}\equiv G M_{500} \mu m_{p}/(2 R_{500})$ and $n_{\mathrm{e
},500}\equiv 500 f_{\rm b} \rho_{\rm c}/(\mu_\mathrm{e} m_{p})$, and $G$ is the gravitational constant, $m_{p}$ is the proton mass, $\mu$ is the mean molecular weight, $\mu_\mathrm{e}$ is the mean molecular weight per free electron, and the critical density $\rho_{\rm c}\equiv 3 H(z)^2/(8\pi G)$ is a function of the Hubble Parameter $H(z)$.  $P/P_{500}$ is plotted as a function of $R/R_{200}$ for the three models in the left panel of Figure \ref{fig:pressure_norm} for three fiducial masses: $L^\star$ halos ($10^{12}\;\msolar$), groups ({``\it Gr''}, $10^{13.5}\;\msolar$), and clusters ({``\it Cl''}, $10^{15}\;\msolar$).  Observations are colored by mass, with three sets of observations shown:  (i) the \citet{Arnaud2010} analytical expression, for three cluster masses; (ii) the \citet{sun2011} median and $1\;\sigma$ spread, for groups with a median mass of $M_{500}=10^{13.85}\;\msolar$;
(iii) the G19 X-COP pressure relationship at $M_{200}=10^{14.9}\;\msolar$, which is overlapped by the DPM cluster profile (and, additionally, Model 1's $L^\star$ and $Gr$ profiles).  We show Model 3 at a range of masses in the right panel of Fig. \ref{fig:pressure_norm} to highlight the progression from convex profiles in clusters to a much flatter power law around galaxies.  We discuss the goodness of the fits with the observationally-derived pressures in \S\ref{sec:obs}.   

\begin{figure*}
\includegraphics[width=0.49\textwidth]{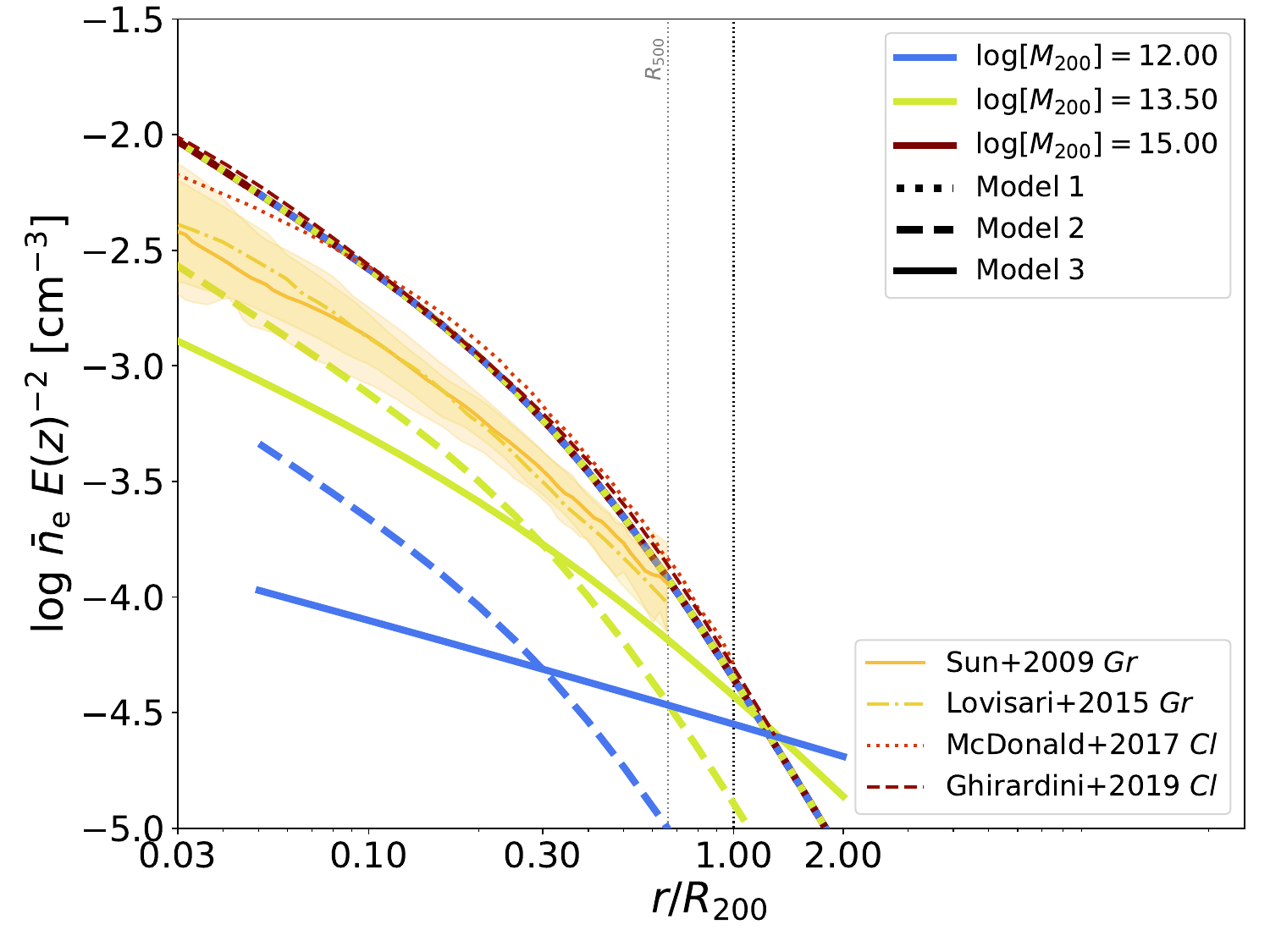}
\includegraphics[width=0.49\textwidth]{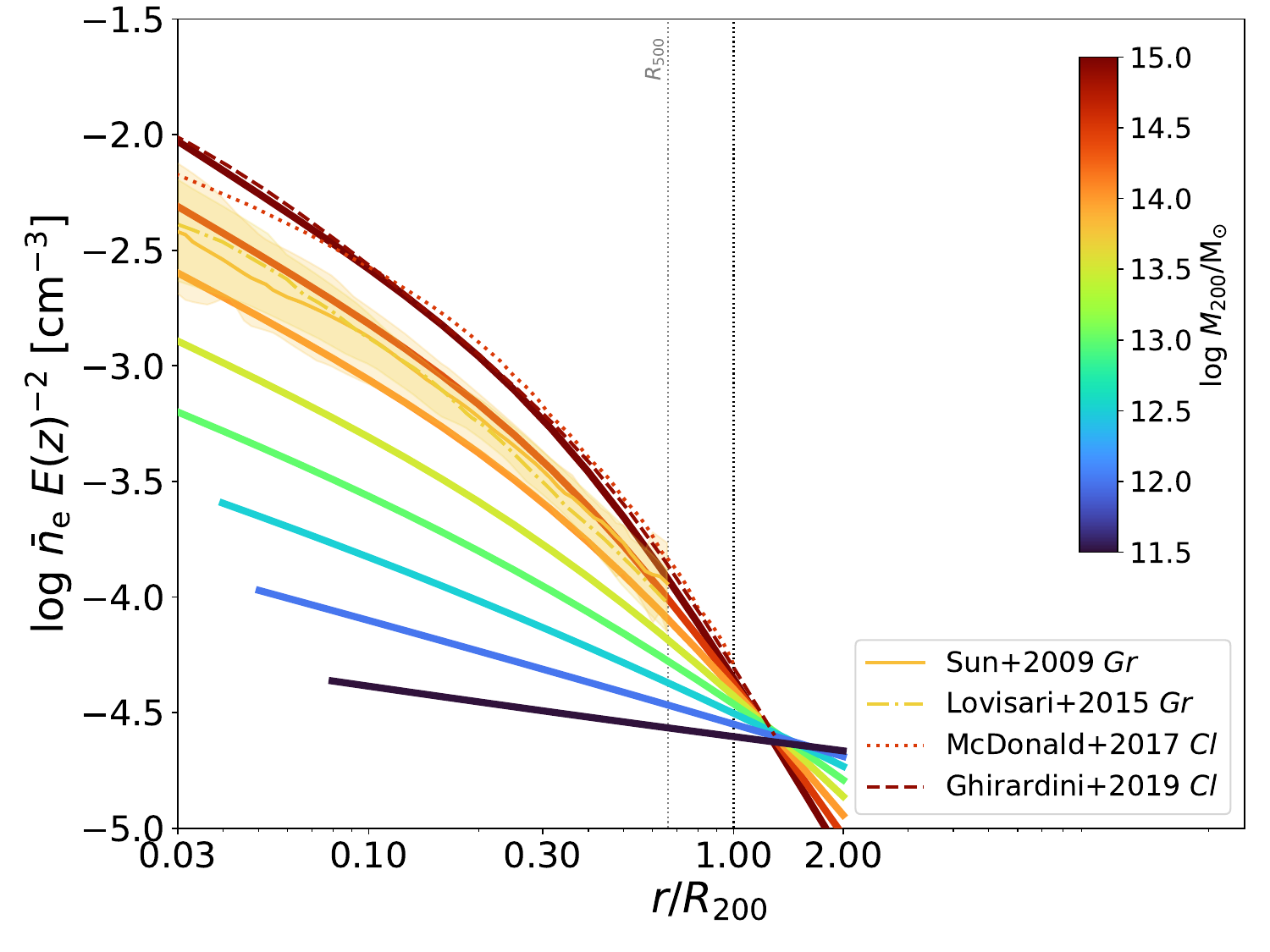}
\caption{{\it Left:} Electron density for the three models at three masses as in Fig.~\ref{fig:pressure_norm}.  {\it Right:} Mass dependence for Model 3, analogous to Fig.~\ref{fig:pressure_norm}.  The \citet{sun2009} and \citet{lovisari2015} groups median profiles, the \citet{McDonald2017} fit to clusters, and the \citet{Ghirardini2019} fit to X-COP clusters are plotted in both panels, with the latter underlying the cluster DPM profiles.}
\label{fig:density}
\end{figure*}
\subsubsection{Electron Density}

Electron density is plotted in Figure \ref{fig:density}.   We show density profiles from \citet{sun2009} (compiled by \citet{Boselli2022}) and \citet{lovisari2015} for groups, and from \cite{McDonald2017} and X-COP (G19) for clusters.  Note that the X-COP profile is overlapped by all Model 1 profiles and the cluster profiles for all three models.   Model 3 shows the same progression for electron densities from clusters to galaxies in the right panel.  Model 3 has a radial slope of approximately $r^{-0.5}$ at $10^{12}\;\msolar$.    

\subsubsection{Metallicity}

\begin{figure}
\includegraphics[width=0.49\textwidth]{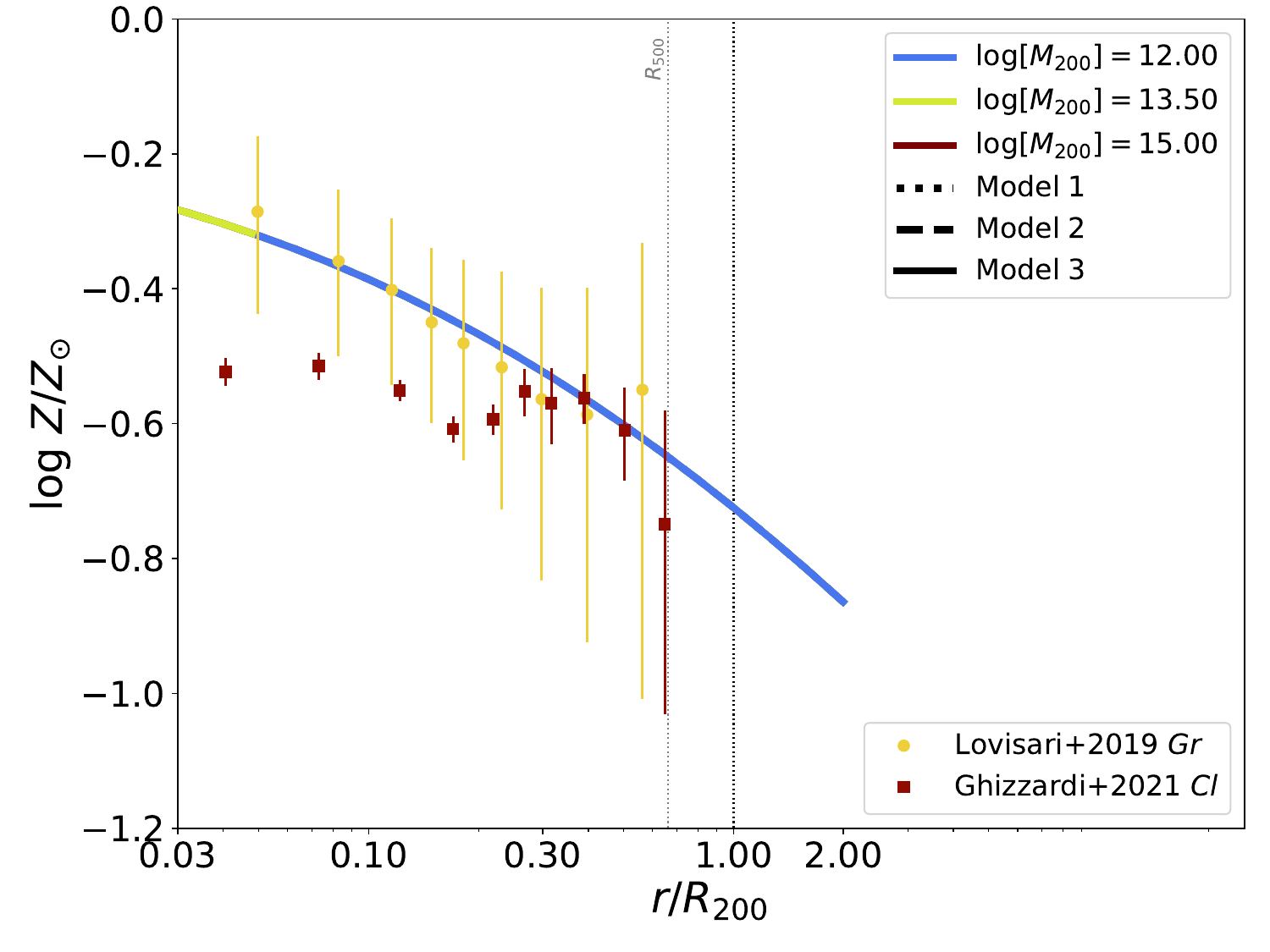}
\caption{Metallicity profiles as a function of radius. All DPM $Z$ profiles are invariant between models, mass bins, and redshifts; hence they overlap.  These profiles are calibrated to the \citet{lovisari2019} groups dataset. } 
\label{fig:metallicity}
\end{figure}

Metallicity ($Z$) profiles normalized by solar metallicity assuming \citet{Asplund2009} abundances are shown in Figure \ref{fig:metallicity}.  We calibrate all our models to the group observations compiled by \citet{lovisari2019}.  The X-COP clusters \citep{Ghizzardi2021} appear to have a flatter profile, but we do not attempt to fit mass-dependent metallicity profiles due to their uncertainty, and leave an exploration of $Z$ variation for the future.  Hence, all DPM metallicity profiles for all masses and models are the same as the profile shown in Fig. \ref{fig:metallicity}.  

\subsection{Secondary Physical Properties} \label{sec:secondphys}

\begin{figure}
\includegraphics[width=0.49\textwidth]{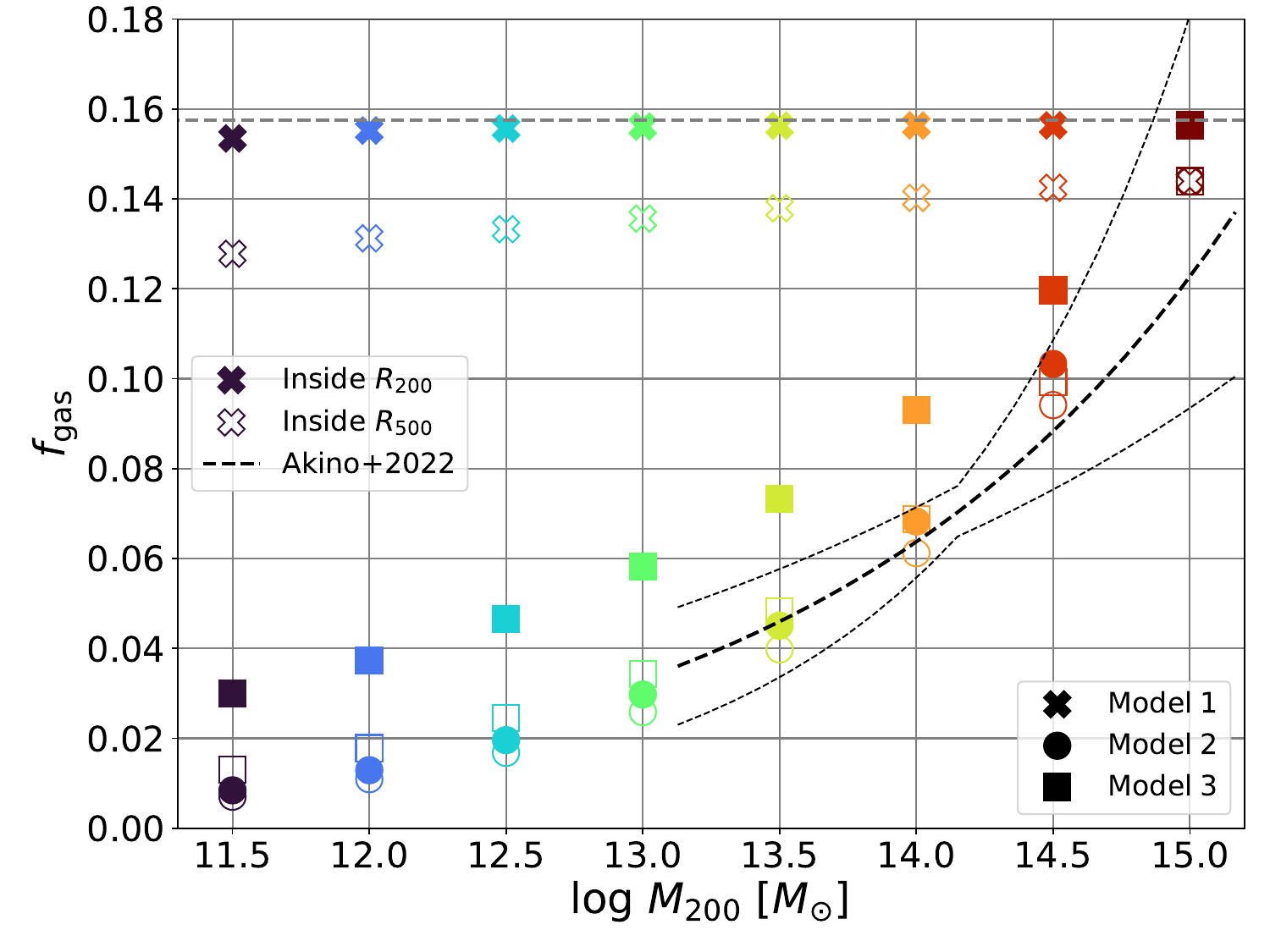}
\caption{Gas fraction as a function of $M_{200}$ inside $R_{200}$ (filled symbols) and inside $R_{500}$ (open symbols).  The dashed horizontal line indicates the cosmic baryon fraction $f_{\rm b}$.  The relationship of \citet{Akino2022} inside $R_{500}$ and the corresponding $1\;\sigma$ range are indicated by black dashed lines.  This relationship should be compared to the open model symbols.  Models 2 and 3 are calibrated to reproduce the \citet{Akino2022} gas fractions.}
\label{fig:fgas}
\end{figure}

Secondary physical properties are synthesized from our primary three physical properties, but they do provide important physical criteria as listed for Model 3 in \S\ref{sec:model3}.  The first physical constraint is atmospheric gas fraction, $f_{\rm gas}$, which we define as $M_{\rm gas}(<R)/M_{\rm total}(<R)$ for a given $R$.  We show two sets of symbols for $R=R_{200}$ and $R_{500}$ in Figure \ref{fig:fgas} for the three models.  The \citet{Akino2022} gas fraction relation inside $R_{500}$ provides a calibration point for Models 2 and 3.

\begin{figure}
\includegraphics[width=0.49\textwidth]{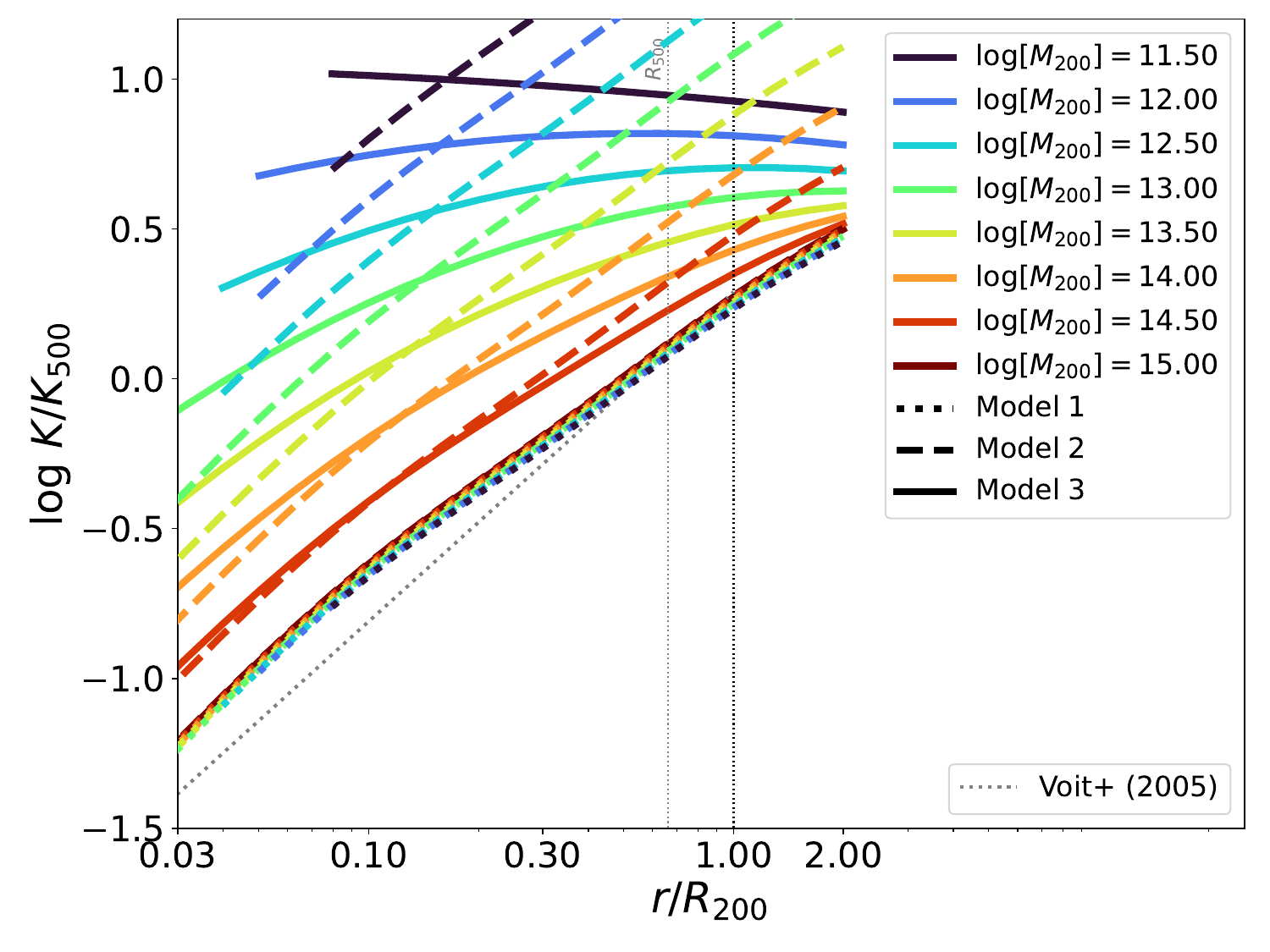}
\caption{Normalized entropy at as a function of radius for the span of halo masses across the three models.  The \citet{Voit2005} baseline entropy profile is plotted as well.  The entropy profiles for Model 1  overlap.}  
\label{fig:entropy}
\end{figure}

We plot astrophysical entropy, $K\equiv T/n_\mathrm{e}^{2/3}$, normalized relative to $K_{500}\equiv T_{500}/n_\mathrm{e,500}^{2/3}$ in Figure \ref{fig:entropy}.  Model 1 is invariant for all halo masses, and follows closely the \citet{Voit2005} baseline entropy profile for clusters without cooling or feedback.  Model 2 maintains the same shape, but shows rising normalized entropy toward lower mass.  Finally, Model 3 indicates increasing and flattening normalized entropy toward low mass.  The tightest criterion listed in \S\ref{sec:model3} is the entropy profile being rising or flat at $L^\star$ halo masses, $\sim 10^{12}\;\msolar$.  In fact, the $10^{11.5}\;\msolar$ halo shows a slightly inverted entropy profile, which may indicate that Model 3 is physically unrealistic and/or potentially dynamically unstable at this mass.   

\begin{figure}
\includegraphics[width=0.49\textwidth]{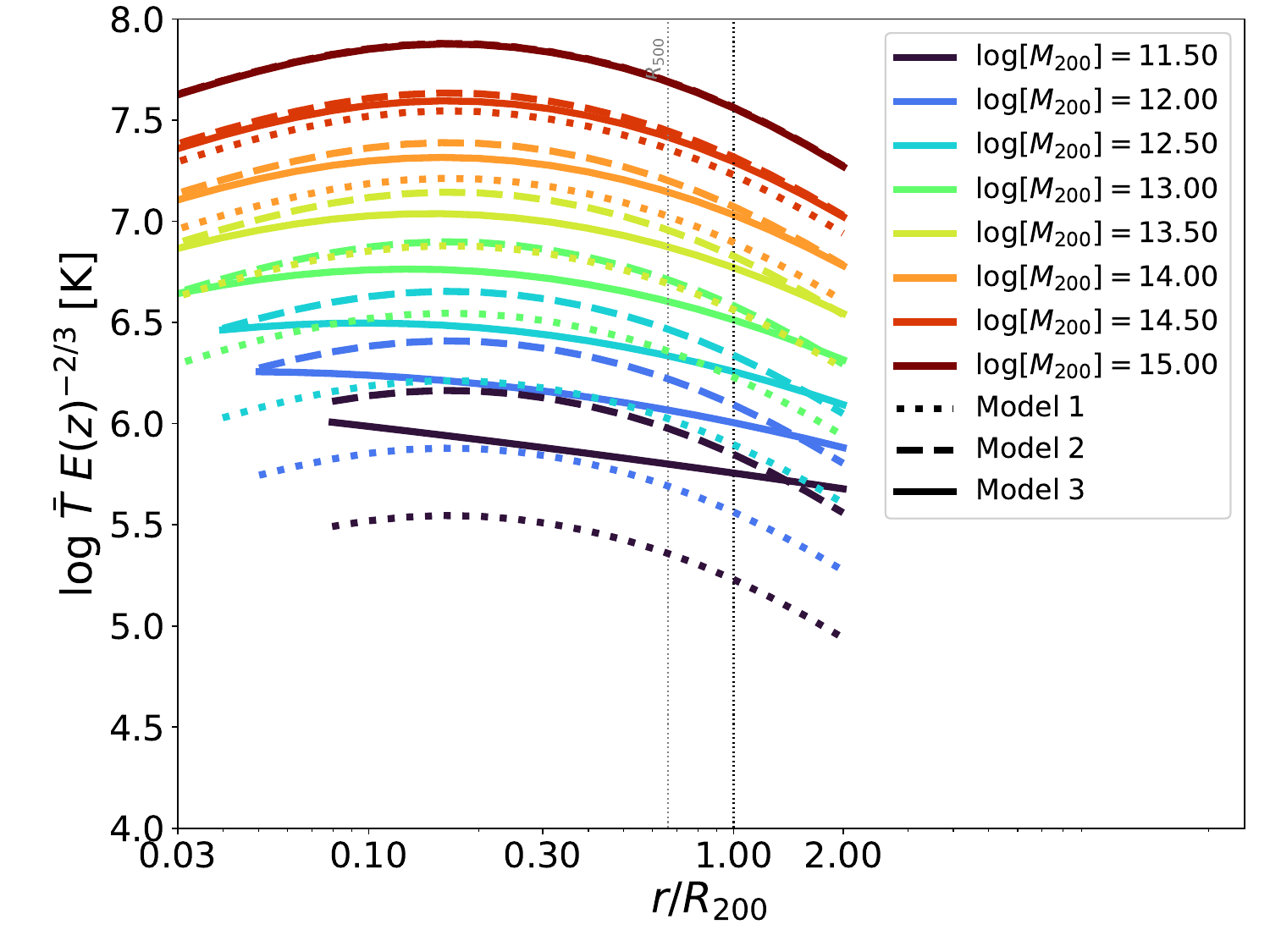}
\caption{Mean temperature profiles of our three models for different halo masses.  Focusing on trends with halo mass, Model 1 has the steepest halo mass temperature dependence, Model 2 has the shallowest, and Model 3 is intermediate. }
\label{fig:temperature}
\end{figure}

We show temperature profiles of the three models at eight halo masses in Figure \ref{fig:temperature}, under the assumption of pure thermal pressure support.  Model 1 is self-similar with temperature scaling according to the $M^{2/3}$ viral relation.  Model 2 is also self-similar, but has a sub-virial temperature scaling of $M^{0.49}$ based on the pressure and density scalings.  Finally, Model 3 also exhibits sub-virial scaling of $M^{0.56}$, but with the added complexity of mass-dependent flattening.  Although  the observations are not plotted here, \citet{lovisari2015} find a mass dependence of $M^{0.6}$ for groups, and \citet{Singh2018} find a much lower slope ($M^{0.3}$ or nearly flat depending on assumptions) for $L^\star$ virial masses ($10^{12}-10^{13}\;\msolar$).

\begin{figure}
\includegraphics[width=0.49\textwidth]{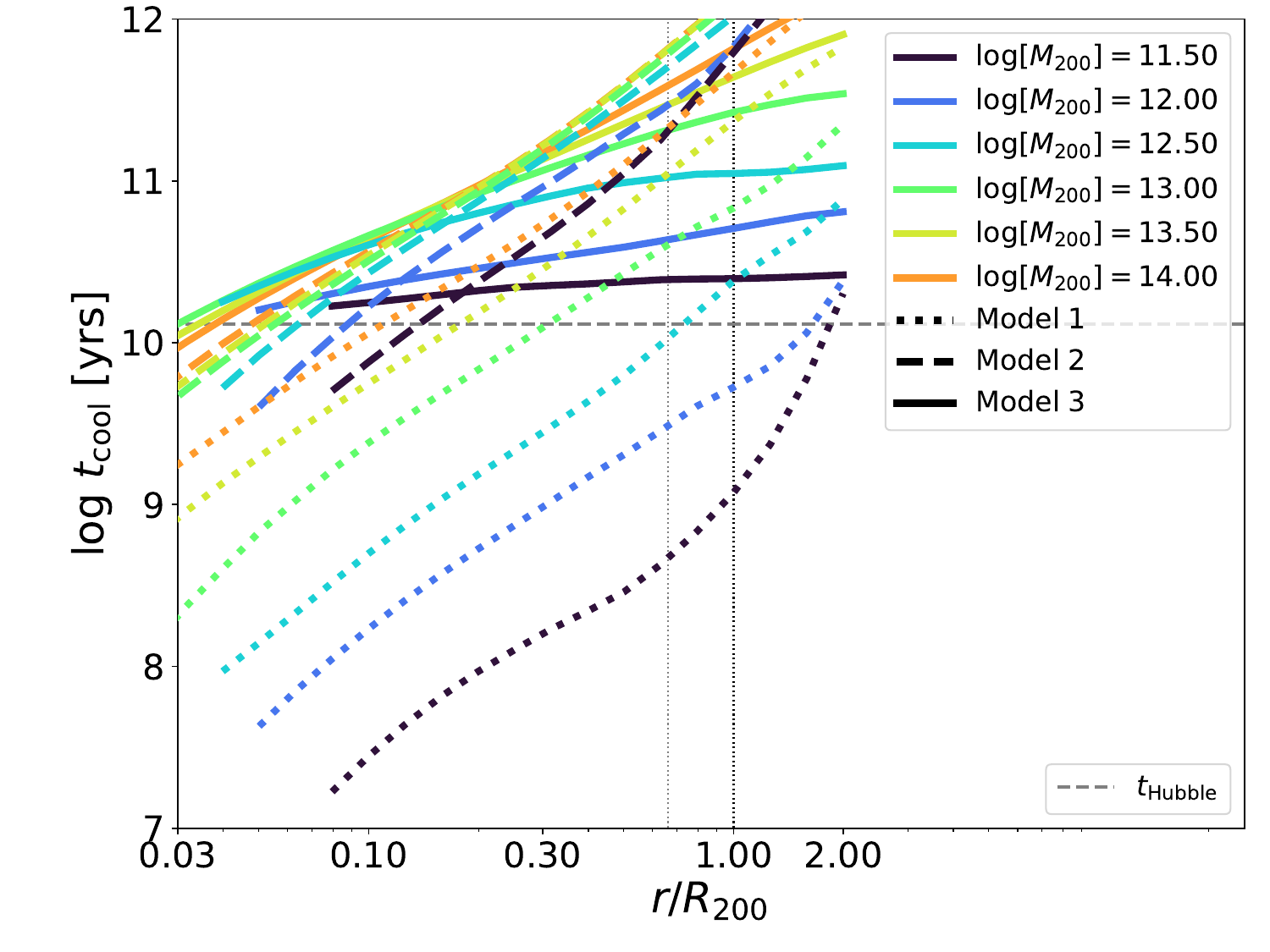}
\caption{Cooling times of $z=0$ $L^\star$ and group halos for the three models.  Cooling times are almost always longer than a Hubble time (horizontal gray dashed line) for Model 3.}
\label{fig:tcool}
\end{figure}

We plot the cooling time profiles for the three models in Figure \ref{fig:tcool}.  Our third criterion for Model 3 is a cooling time comparable to or longer than the Hubble time at radii beyond the extent of the galaxy.  These cooling times assume isobaric cooling, which have $5/3\times$ longer cooling times than isochoric cooling due to $P dV$ work.  Model 1 has much shorter cooling times for $L^\star$ halos, which is another reason to cast doubt on this model.  

\section{Comparison to Observations}\label{sec:obs}

We now compare our DPMs to observational datasets arranged by waveband: X-ray (\S\ref{sec:obsXray}), sub-mm/mm (\S\ref{sec:obsSZ}), radio (\S\ref{sec:obsDM}), and UV (\S\ref{sec:obsUV}).  Within each subsection, we discuss the comparison along with potential observational biases.  Our approach contrasts with other papers that primarily use $L^\star$ CGM datasets to constrain the models \citep[e.g.][]{Faerman2020, Stern2024, Oren2024, Singh2024}.  We use physical property profiles anchored at the cluster and group scale to extrapolate our three models to the CGM regime.  Our atmospheric models will sometimes underestimate an observable, which we argue may not invalidate that model because the non-spherical, often ``cloud''-like, component\footnote{This component refers to anything non-spherical with respect to the halo centre and can include shells of X-ray emitting gas, for example.  Wind-blown bubbles could lead to a deficit compared to the atmospheric signal, but we assume the non-spherical component is usually an additive component.} is not included.  We default to $\sigma^{n_\mathrm{e}}=0.15$ unless specified otherwise.  Varying $\sigma^{n_\mathrm{e}}$ only affects X-ray and UV observations, because pressures and densities obtained from SZ Effects and electron DMs are invariant by construction of the DPM. 

\subsection{X-ray Observables} \label{sec:obsXray} 

\subsubsection{{\it eRASS:4 Survey}} \label{sec:obseRASS4} 

\begin{figure}
\includegraphics[width=0.49\textwidth]{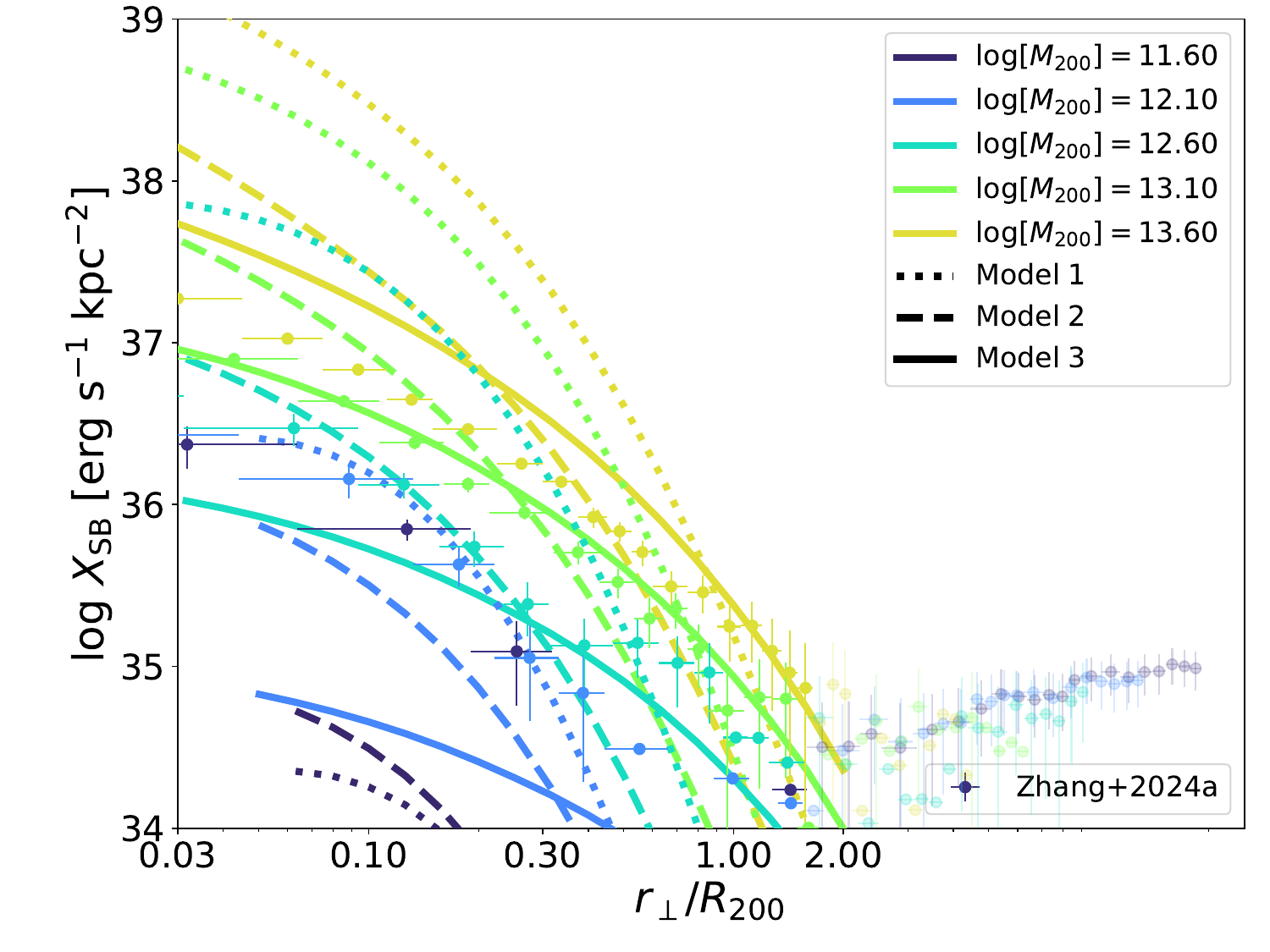}
\caption{Soft X-ray surface brightness profiles for the three models overlaid with {\it eROSITA} eRASS:4 datapoints from \citet{Zhang2024a} as a function of fractional projected radius.  DPM profiles spanning $M_{200}=10^{11.6-13.6}\;\msolar$ are plotted, corresponding to the 5 sets of DPM profiles.  Transparent datapoints that are beyond $R_{200,m}$ are not used for statistical comparisons.   }
\label{fig:SoftXray}
\end{figure}

Soft X-ray surface brightness ($X_{\rm SB}$) profiles measured from stacked X-ray emission with the {\it eROSITA} eRASS:4 Survey are plotted for $L^\star$ halos and groups in Figure \ref{fig:SoftXray} \citep{Zhang2024a}.  The five halo mass bins correspond to log($M_{200,m}/\msolar)=11.5-12.0,\;12.0-12.5,\;12.5-13.0\;13.0-13.5$\;\&\;$13.5-14.0$, where each represents the stacked CGM emission from the CEN$_{\rm halo}$ sample of \citet[][see their fig. 6]{Zhang2024a}\footnote{We convert $M_{200,m}$ to $M_{200}$.}.

\paragraph{Observational Comparison and Biases} 

\begin{figure}
\includegraphics[width=0.49\textwidth]{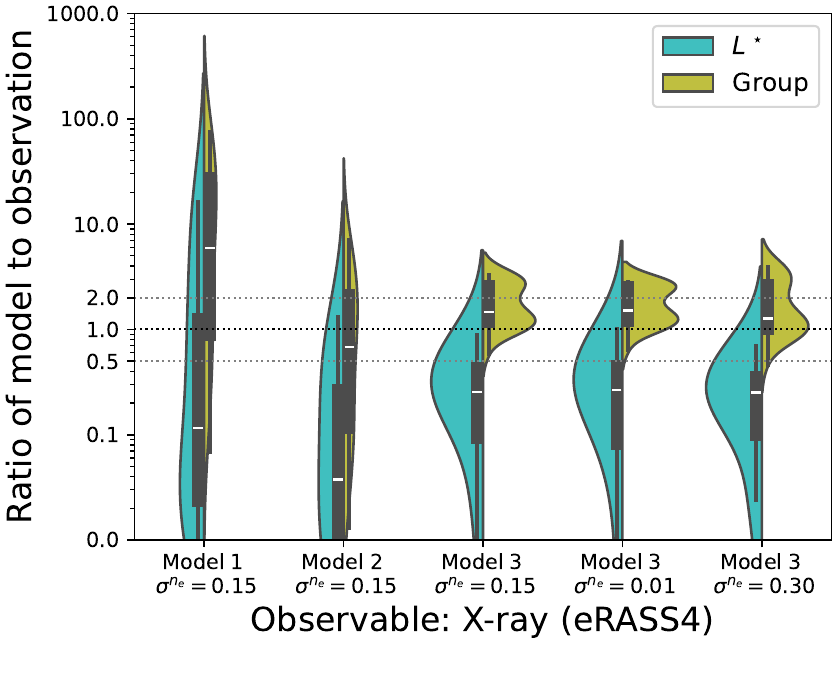}
\caption{Violin plot plotting the ratio of the DPM soft X-ray surface brightness profiles $X_{\rm SB}$ over the \citet{Zhang2024a} eRASS:4 datapoints in Fig. \ref{fig:SoftXray}.  Two sets of model over observation ratios correspond to the three $L^\star$ stacks (teal) and two group stacks (dark yellow).  Model 1 has the highest ratios indicating significant overestimates at low radii especially for groups, as well as low ratios indicating underestimates at high radii and for $L^\star$ halos (cf. Fig. \ref{fig:SoftXray}).  Model 2 has lower overall $X_{\rm SB}$ and a large range of ratio of model to observation values, which is indicative of DPM profile shapes being steeper than observations.  Model 3 has weaker X-ray emission, while the profile shapes are more similar to observed as reflected by the smaller spreads, especially for groups, while $L^\star$ halos are under-luminous.  The density dispersion ($\sigma^{n_e}$) has little effect on $X_{\rm SB}$ as reflected by the similarity of the latter three ratio ranges. }
\label{fig:violin_eRASS4}
\end{figure}

We introduce the use of violin plots to show the ratio of DPM predictions relative to observed datapoints from \citet{Zhang2024a} in Figure \ref{fig:violin_eRASS4}.  These plots compile the distributions of DPM predictions given the radial position, halo mass, and redshift of an observed datapoint and compare them to the datapoints.  We plot the ratios of the model with published datapoints, independent of errors, to show how the DPM profiles perform relative to available data.  In this case, we take the ratios of our profiles and $X_{\rm SB}$ datapoints from five observed stacked halo mass bins inside $R_{200,m}$ shown in Fig. \ref{fig:SoftXray} as an evaluation of DPM performance.  We show separate distributions for the three $L^\star$ and two group stacks.  Model 1 shows a large spread, including cases where the DPM exceeds the observed data by a factor of 10$\times$ (and occasionally even 100$\times$). These extreme overestimates, arising at the smallest radii (Fig. \ref{fig:SoftXray}), confirm that $L^\star$ halos cannot be scaled down analogues of clusters, as has been known since their non-detection with {\it ROSAT} \citep{benson2000} and recently quantified using simulations \citep[e.g.][]{lau2025}.  The large statistical spread shows that the DPM profiles are much steeper than the eRASS:4 datapoints.  Model 2 is lower than Model 1, except at the lowest masses owing to higher temperatures for Model 2.  

Model 3 has a median ratio of nearly unity across all five stacks, as well as a much smaller spread overall, signifying that it is the best performing model.   The $10^{13.1}\;\msolar$ profile of Fig.~\ref{fig:SoftXray} nearly overlaps the datapoints, while lower mass profiles progressively underpredict $X_{\rm SB}$ compared to observations, especially at small radii.  We prefer our models to underestimate observations due to our models not containing a non-atmospheric component that can also create X-ray emission, which we argue is likely significant due to the non-spherical emission observed around $L^\star$ galaxies, as discussed in \S\ref{sec:mockXray}.  However, the highest mass bin probing the group regime overestimates the datapoints, but this overestimate is still smaller than for the other models.  

Density/temperature dispersion has at most a $0.1$ dex effect for X-ray emission, with larger dispersions slightly lowering emission for $L^\star$ halos, leaving intermediate group halos nearly unaffected and slightly raising emission for cluster profiles (not shown).  The range of masses included in Fig. \ref{fig:violin_eRASS4} washes these differences out making the last three ranges nearly indistinguishable.  

\subsubsection{{\it X-ray-based Group Profiles}} \label{sec:obsXraygroups} 
 
We have already included the \citet{sun2011} sample pressure profile with $M_{200}\sim 10^{14.0}\;\msolar$ in Fig. \ref{fig:pressure_norm} (also shown in Fig. \ref{fig:pressure_all}) and the \citet{sun2009} and  \citet{lovisari2015} density profiles at $M_{200}\sim 10^{13.7}-10^{13.8}\;\msolar$ in Fig. \ref{fig:density}.  

\paragraph{Observational Comparison and Biases} 

Model 1's pressure is 30\% higher and its density is $2\times$ higher than the observationally-derived data.  Model 2 has pressures and densities that are $\sim 80\%$ and $\sim 70-80\%$ the data respectively, which is the best performance of all DPMs.  Model 3 has pressure and density being $\sim 50\%$ and $\sim 60-75\%$ the data respectively.  

These groups, primarily observed based on detection in the {\it ROSAT} survey, are X-ray-selected and are likely biased, which is further supported by the $f_{\rm gas}$ for groups being higher in the \citet{lovisari2015} sample compared to that of \citet{Akino2022}.  The latter sample uses weak lensing and multi-band photometry to correct for X-ray selection biases, and this leads to lower values for physical properties derived from X-ray-observed groups.  The X-GAP ({\it XMM} Groups AGN Project) survey \citep{Eckert2024} has observed a sample to overcome X-ray selection bias of groups, but pressure and density profiles from this survey have not yet been published.  

\subsubsection{X-ray Luminosity Scaling Relations} \label{sec:obsXrayscaling}

\begin{figure}
\includegraphics[width=0.49\textwidth]{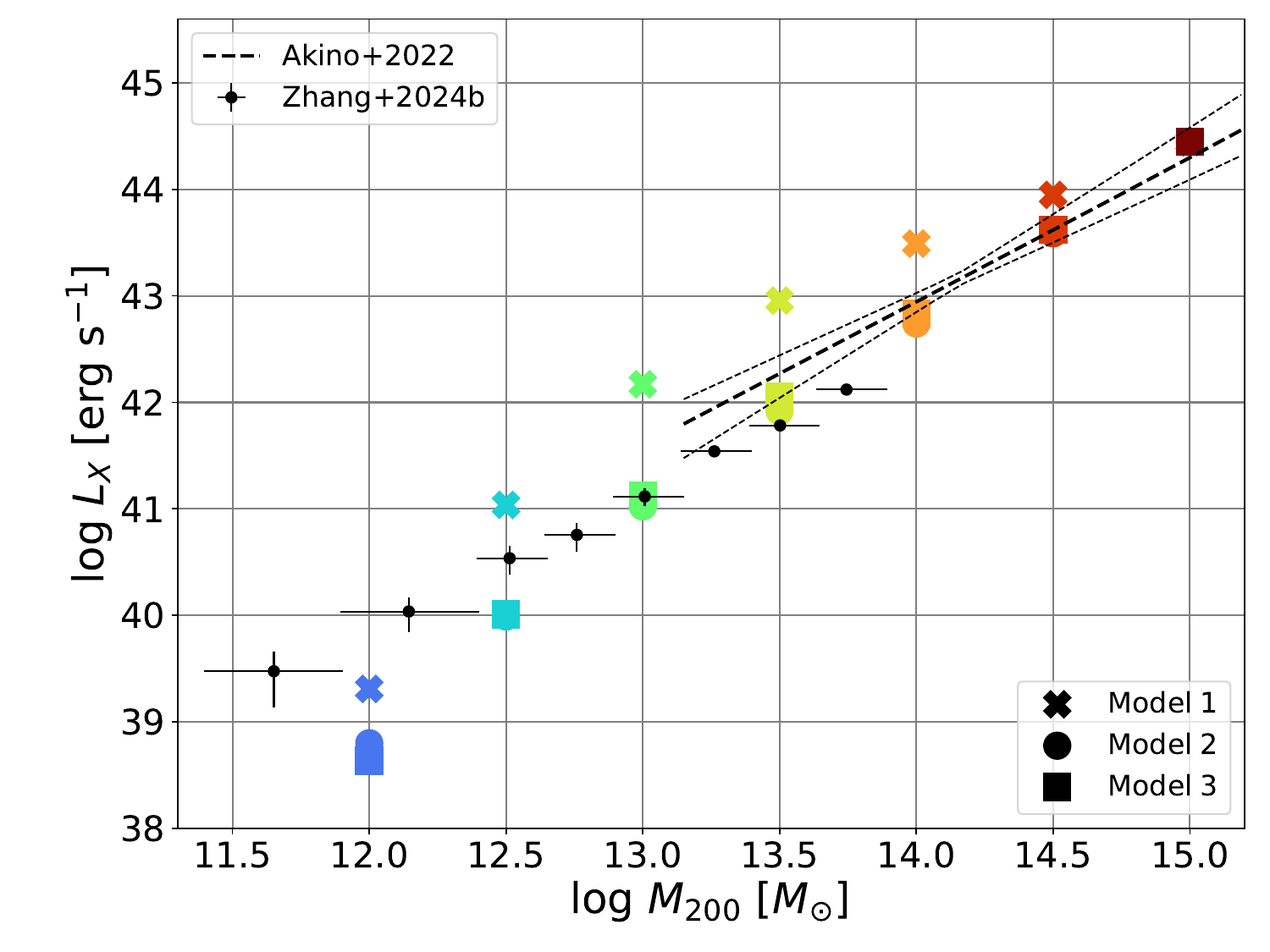}
\caption{The $L_{X}-M_{200}$ relationship for emission summed between $0.15-1.0 R_{500}$.  The black dashed line indicates data from groups and clusters compiled by \citet{Akino2022}. Black points represent data from \citet{Zhang2024b} based on {\it eROSITA} eRASS:4 stacked samples.}
\label{fig:LX}
\end{figure}

We plot X-ray luminosities summing 0.5-2.0 keV emission between $0.15-1.0$ $R_{500}$ in our DPMs in Figure \ref{fig:LX}.  We overplot the \citet{Akino2022} relationship for groups/clusters and the binned data points from \citet{Zhang2024b}, who summed within $1.0$ $R_{500}$.  Slopes, defined as $L_X \propto M^{\beta}$, are $\beta=1.38$ and $1.3$ for groups/clusters and galaxy halos, respectively.  The scaling relations of groups and clusters are well studied, and the \citet{Lovisari2021} review of scaling relations finds slopes between $1.07-1.66$ across different studies.  

\paragraph{Observational Comparison and Biases} 

For Model 3, we find a slope $\beta = 1.5$ for groups/clusters and a much steeper $\beta=2.3$ for halos between $10^{12.0-13.5}\;\msolar$.  The reasonable agreement for groups/clusters is expected, since these are anchors for Model 3.  The decline is faster for $L^\star$ halos, which also agrees with the trend in Fig~\ref{fig:SoftXray}.  Again, the underestimate relative to eRASS:4 data at progressively lower mass allows the possibility for X-ray emission to arise from non-atmospheric CGM structures as discussed in \S\ref{sec:obseRASS4}.  

While the halo mass in \citet{Akino2022} is directly measured from weak lensing observations, the halo mass in \citet{Zhang2024a, Zhang2024b} is assigned through the group finder from \citet{Tinker2021} and calibrated to agree with the SMHM conversion from \citet{Mandelbaum2016}. However, there still exists scatter between SMHMs measured in
different works through weak lensing \citep{Mandelbaum2016, Bilicki2021}. The inconsistency between \citet{Akino2022} and \citet{Zhang2024b} where they overlap in the group regime might arise from the difference in $M_{\rm halo}$.  Upcoming galaxy surveys and weak lensing studies will improve the
accuracy of SMHM.  

\subsection{Sunyaev-Zeldovich Observables} \label{sec:obsSZ} 

\subsubsection{$z\sim 0.0$ tSZ Stacks} \label{sec:obstSZ}

\begin{figure}
\includegraphics[width=0.49\textwidth]{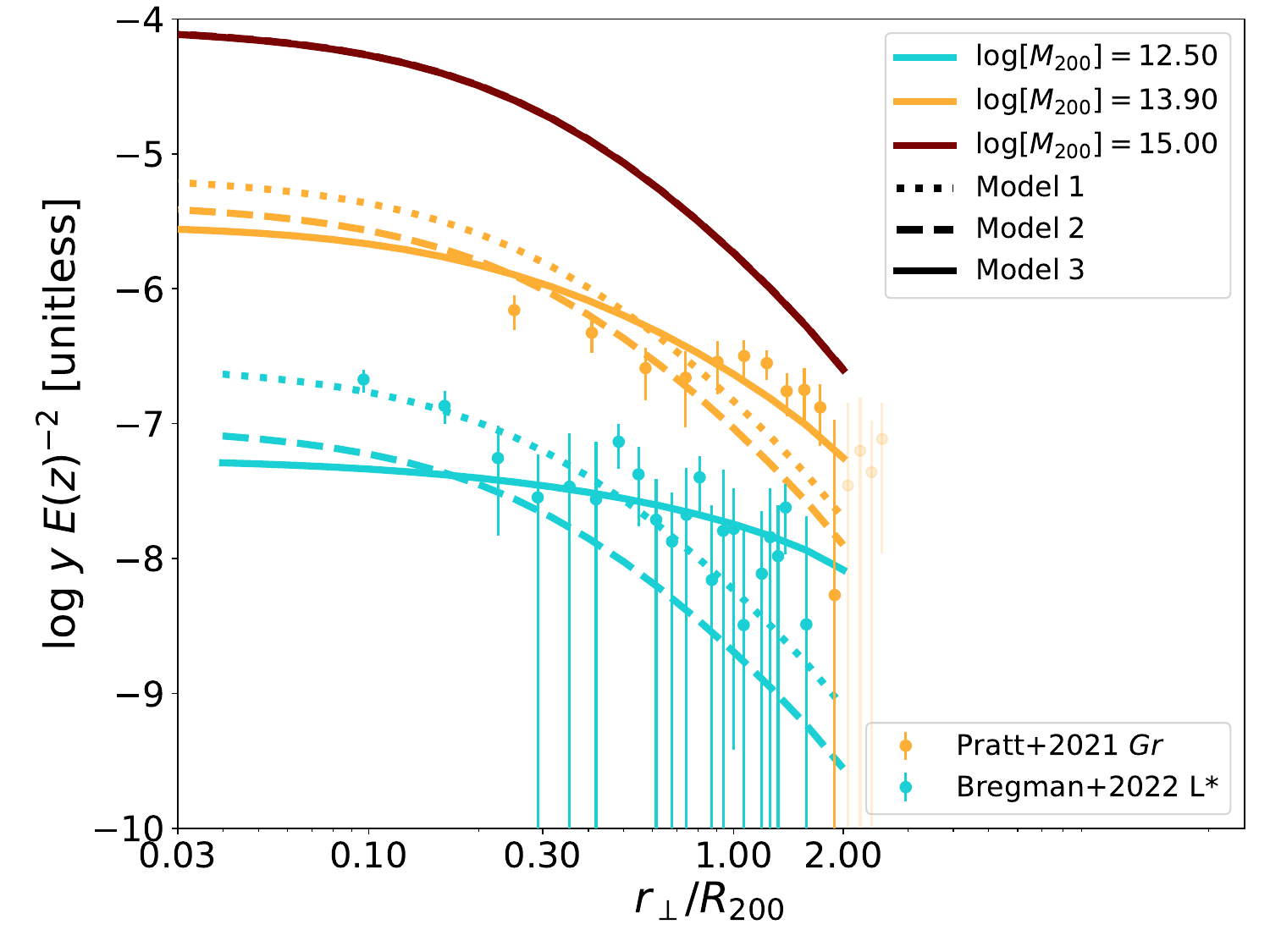}
\includegraphics[width=0.49\textwidth]{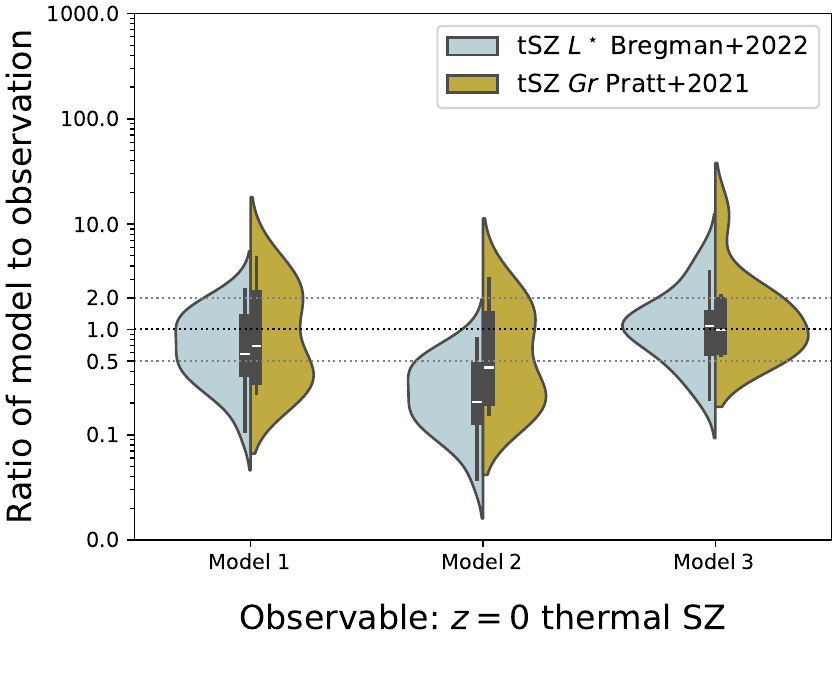}
\caption{The upper panel shows the Compton $y$ parameter from stacked thermal SZ (tSZ) profiles at $z=0.0$.  Observational datasets are from \citet{bregman2022} for $L^\star$ halos (cyan), and from \citet{pratt2021} for group halos (orange). For each of the three DPM variants, we show the two halo masses corresponding to these data as correspondingly coloured lines, in addition to clusters (dark red).  The lower panel shows a violin plot comparing the DPM SZ predictions to these observed datasets.  }
\label{fig:ySZ}
\end{figure}

In Figure \ref{fig:ySZ} (upper panel), we plot the Compton $y$ parameter for the \citet{pratt2021} group sample averaging $M_{200}=10^{13.9}\;\msolar$ and the \citet{bregman2022,bregman2024} $L^\star$ spiral sample\footnote{\citet{bregman2024} is an Erratum where they revised $y$ measures upward by $1.9\times$ compared to \citet{bregman2022}.} with an average $M_\star=10^{10.8}\;\msolar$, which we convert to a halo mass of $M_{200}=10^{12.5}\;\msolar$ via abundance matching.  The orange and cyan model DPM profiles correspond to these halo masses, and we additionally show the cluster profiles (dark red).  The $y$ parameter is proportional to projected pressure.  The comparison is performed at $z=0.0$.  

\paragraph{Observational Comparison and Biases} 

There are only 11 $L^\star$ halos and 10 groups in this comparison.  The weighting of halos goes as $M_{200}^{5/3}$ if they follow virial relationships, which means even one underestimated halo mass could significantly bias the results.  We show the violin plots of the datapoints and DPM profiles in the lower panel of Fig. \ref{fig:ySZ}.  

For $L^\star$ halos \citep{bregman2022, bregman2024}, Model 3 performs the best followed by Model 1 with Model 2 being the worst fit.  However, we note that this comparison weights all datapoints equally even if they have large error bars, and the comparison is driven by datapoints at larger radii.  Model 3 performs worst for the two innermost datapoints.  For groups \citep{pratt2021}, Model 3 performs best with the median ratio of model to observation being near unity; however the spread in values reflects that the datapoints have a flatter profile than Model 3 in Fig. \ref{fig:ySZ}.  

\subsubsection{BOSS CMASS $z=0.55$ tSZ \& kSZ Stacks} \label{sec:obsCMASS}

\begin{figure}
\includegraphics[width=0.49\textwidth]{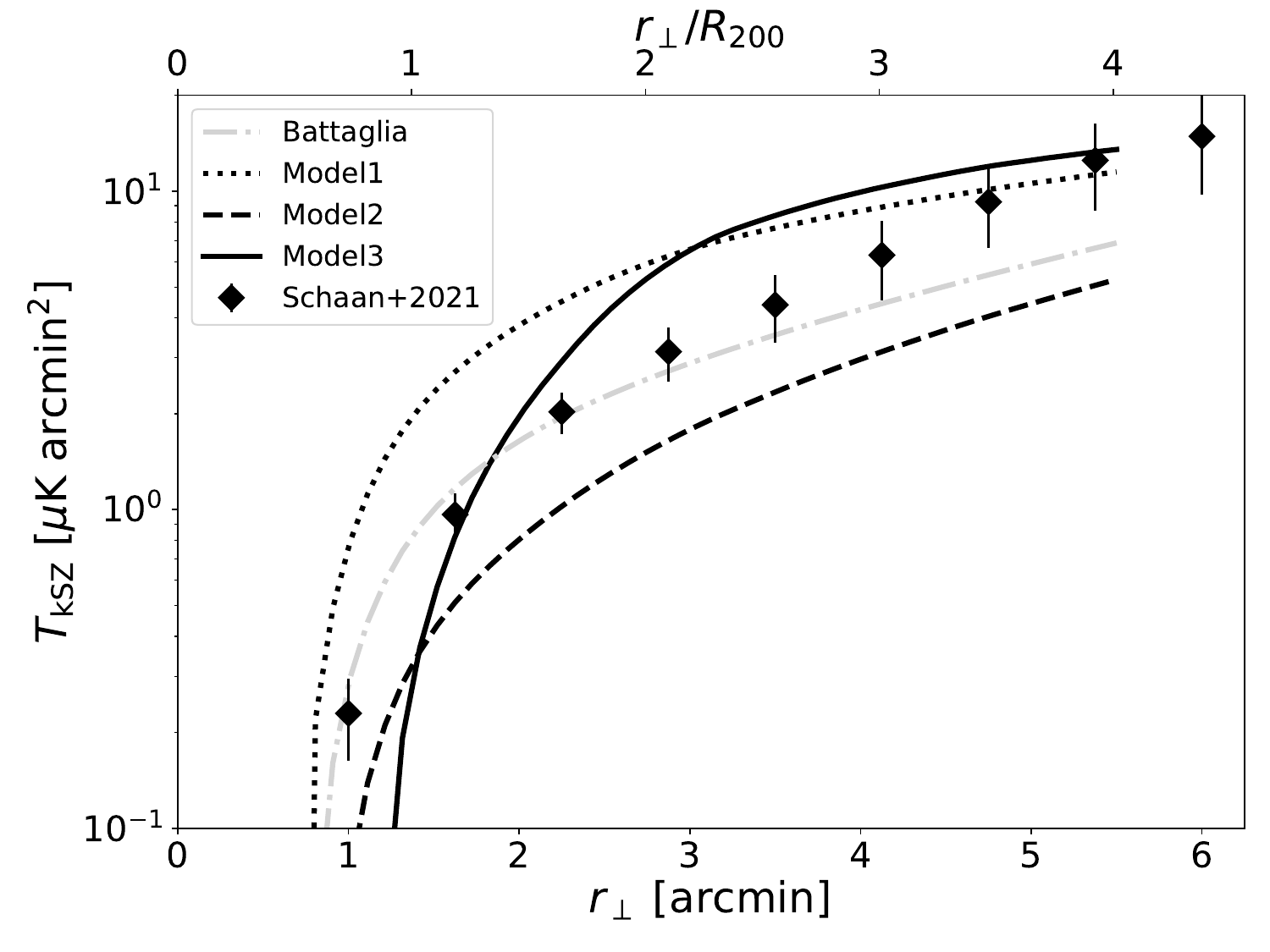}
\includegraphics[width=0.49\textwidth]{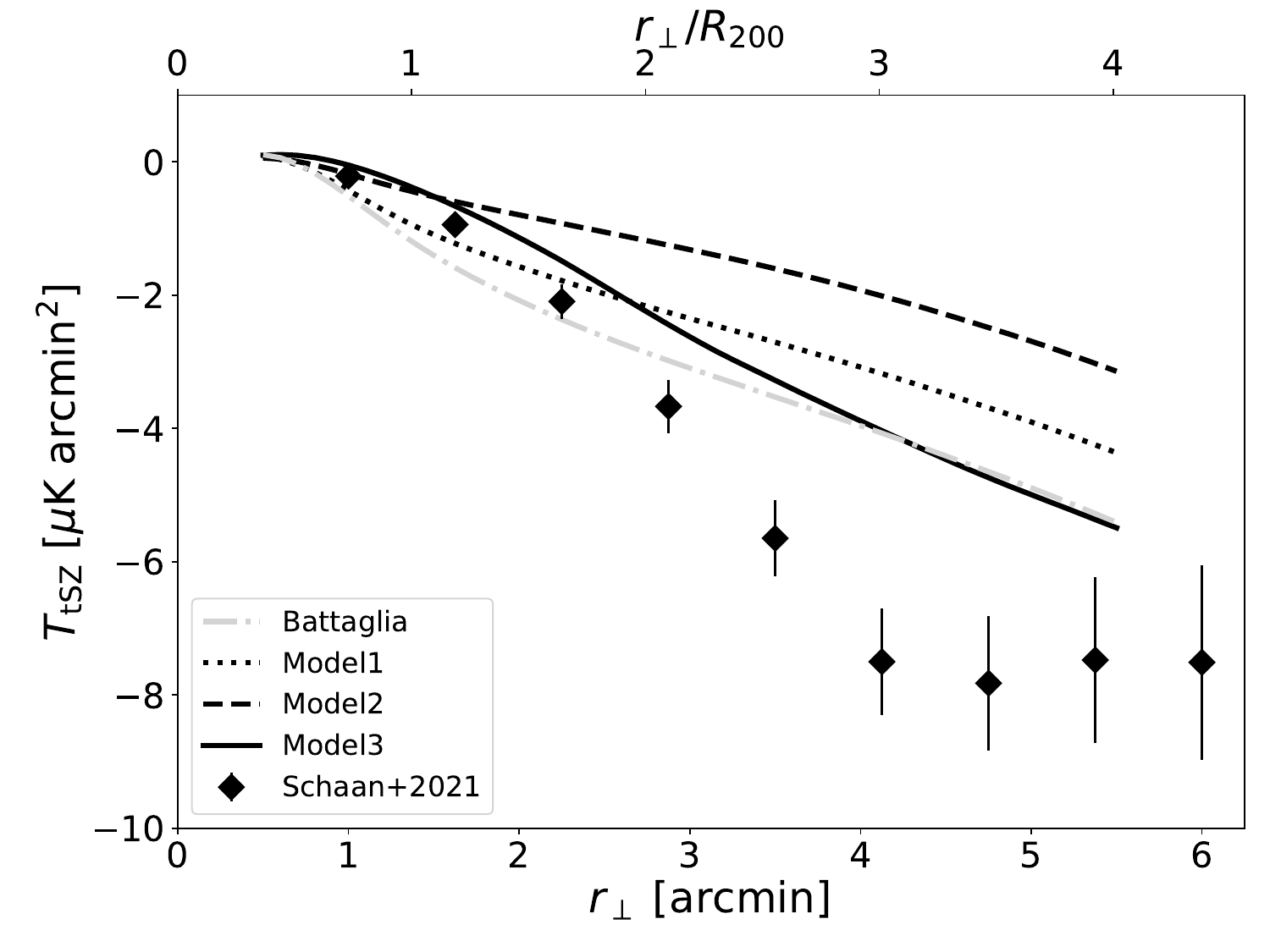}
\includegraphics[width=0.49\textwidth]{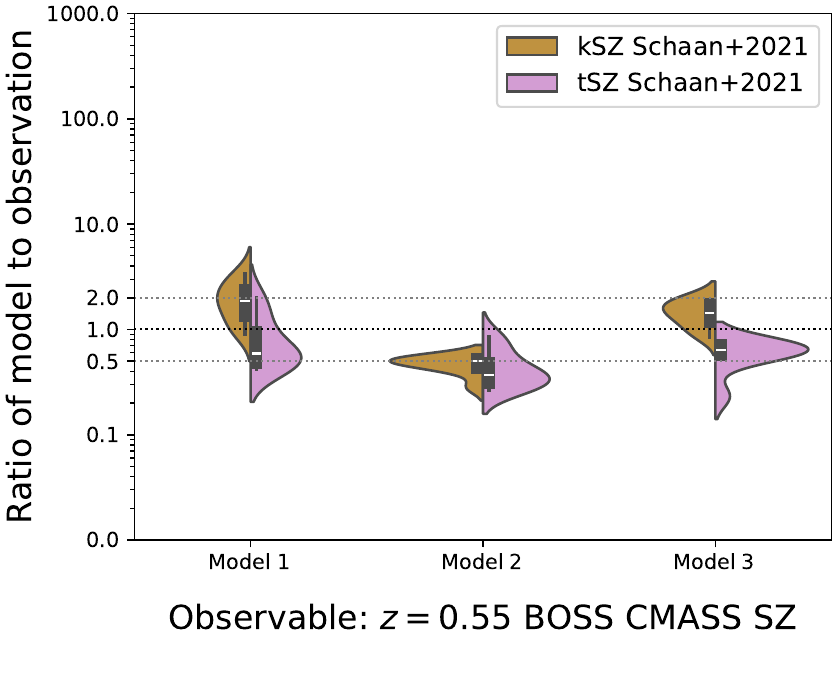}
\caption{The upper panel shows cumulative kinetic SZ signal from the DPM profiles and the \citet{Schaan2021} observational datapoints for halos averaging $M_{200}=10^{13.5}\;\msolar$.  The middle panel shows the cumulative thermal SZ decrement for these same datasets.  The observed stacks are 150 GHz dust-corrected, and the DPM profiles are modelled using the {\tt Mop-c-GT} package.  The gray dash-dot profiles are from \citet{battaglia2016} for kSZ and \citet{battaglia2012} for tSZ.  The lower panel shows the violin plot of the three models relative to the \citet{Schaan2021} data, showing discrepancies in all cases.   }
\label{fig:SZ_Schaan21}
\end{figure}

The \citet{Schaan2021} datapoints from stacked CMASS halos averaging $M_{200}=10^{13.5}\;\msolar$ and $z=0.55$ are plotted in Figure \ref{fig:SZ_Schaan21}.  This dataset uses $\sim 4\times 10^5$ galaxies defined by having ``constant mass'' (CMASS) to measure the SZ temperature distortion in the ACT CMB map.  The kSZ dataset in the upper panel is the 150 GHz dataset shown in the upper panel of fig. 7 of \citet{Schaan2021}, and the tSZ dataset in the middle panel is the 150 GHz corrected for dust shown in the upper panel of fig. 8 of the same paper.  We input our profiles into the {\tt Mop-c-GT} software using the four halo mass bins and their fractional contributions to the entire stack as specified within the software.  The lines represent the result of this modelling for our three models, as well as the profiles derived from earlier results of \citet{battaglia2012} and \citet{battaglia2016} shown as dash-dot gray lines.  We note that these latter profiles were explored in \citet[][see their fig. 6 lower panels]{Amodeo2021,Amodeo2023}, and are shown because they produce the same result as that paper\footnote{The Battaglia kSZ signal was revised upward in the Erratum of \citet{Amodeo2023} and appears similar to the gray dash-dot line we calculate using {\tt Mop-c-GT}.} allowing a relative comparison to our DPM profiles.  

\paragraph{Observational Comparison and Biases} 

No DPM profiles reproduce the $T_{\rm kSZ}$ nor $T_{\rm tSZ}$ signals in Fig. \ref{fig:SZ_Schaan21}, which we quantify in a violin plot in the lower panel of this figure.  For kSZ, Model 1 overpredicts and Model 2 underpredicts the ionized baryon content at most radii, while the \citet{battaglia2016} profile with an intermediate baryon content falls closest to observations in Fig. \ref{fig:SZ_Schaan21} (upper panel).  Model 3's flatter density profiles, resulting in a steeper cumulative $T_{\rm kSZ}$ profile, are unsupported by this dataset.  

All models underpredict the tSZ signal, especially at radii $>2 R_{200}$ indicating that the measured pressures are higher than DPM profiles.  \citet{Amodeo2021} also found that the IllustrisTNG simulations underpredicted both kSZ and tSZ signals, and \citet{Moser2022} found that none of the CAMEL Simulations varying a large range of feedback parameters could reproduce the strength of the CMASS tSZ signal, even considering possible systematics \citep{Moser2023}.  

The ratio of tSZ over kSZ can constrain the temperature, as in \citet{Schaan2021}.  We note that Model 1 has the lowest ratio and Model 2 has the highest ratio, which agrees with these models having the lowest and highest temperatures, respectively; Model 3 is intermediate.  The violin plots reflect that none of our three DPM cases fits the \citet{Schaan2021} datasets, and they additionally show that comparison of the pink (tSZ) and cyan (kSZ) distributions indicate these temperature trends.  

\subsection{Electron Dispersion Observables} \label{sec:obsDM}

\begin{figure}
\includegraphics[width=0.49\textwidth]{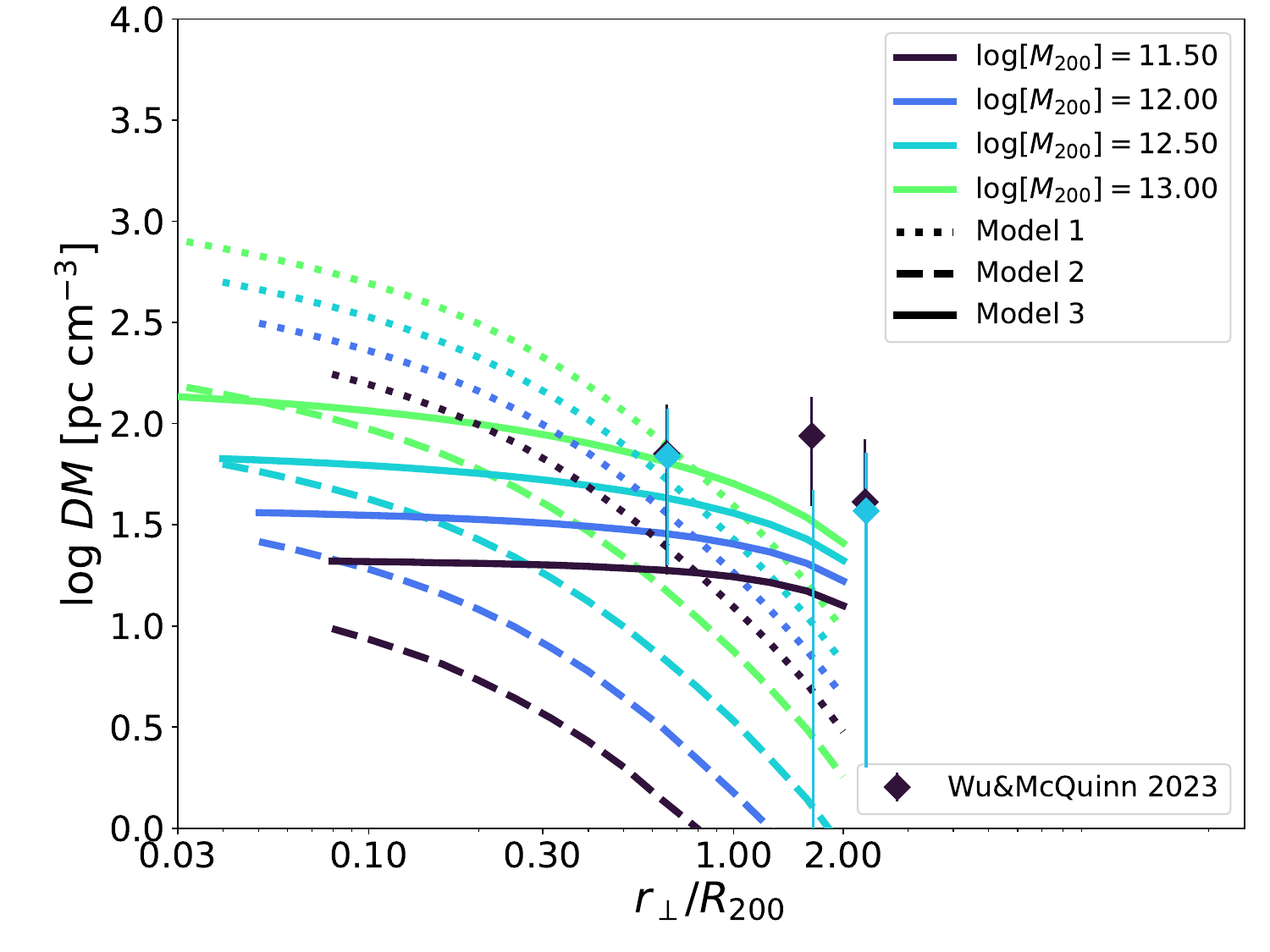}
\caption{Dispersion measures from the three DPMs at $L^\star$ halos masses, compared to datapoints collected by \citet{Wu2023}.  The large errorbars indicate that this observable is not yet able to constrain models. 
}
\label{fig:DM}
\end{figure}

The next observable is electron DMs as a function of impact parameter, constrained in two halo mass bins at three radii by \citet{Wu2023}.  We plot $L^\star$ $z=0.0$ DM profiles in Figure \ref{fig:DM} along with these data.  

\paragraph{Observational Comparison and Biases} 

The comparison of DPM profiles to \citet{Wu2023} datapoints is unconstraining at its current level. Although it appears that Model 2 underpredicts the data, it is possible that this is because ionized cool CGM clouds are a significant contributor to the DM measured around $L^\star$ galaxies.  We highlight that Models 2 and 3 have highly contrasting predictions at large radii that future surveys should soon be able to test.  Interestingly, while $L_X$ values are similar for these two models in the $L^\star$ regime (cf. Fig. \ref{fig:LX}), DM profiles provide an orthogonal constraint. 

\subsection{Ultraviolet Observables} \label{sec:obsUV} 

\subsubsection{$\OVI$ Absorption} \label{sec:obsOVI} 

\begin{figure*}
\includegraphics[width=0.67\textwidth]{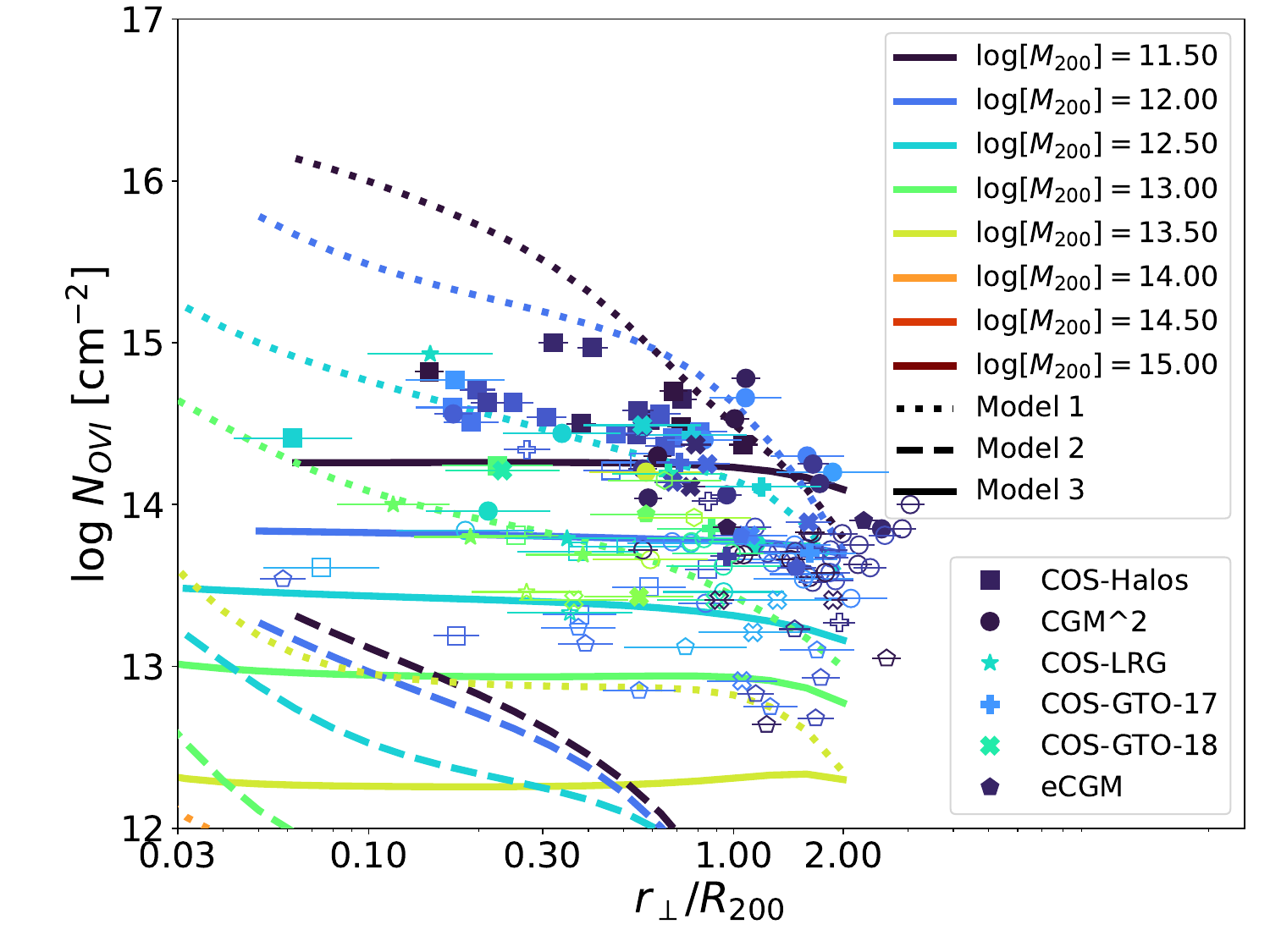}
\includegraphics[width=0.49\textwidth]{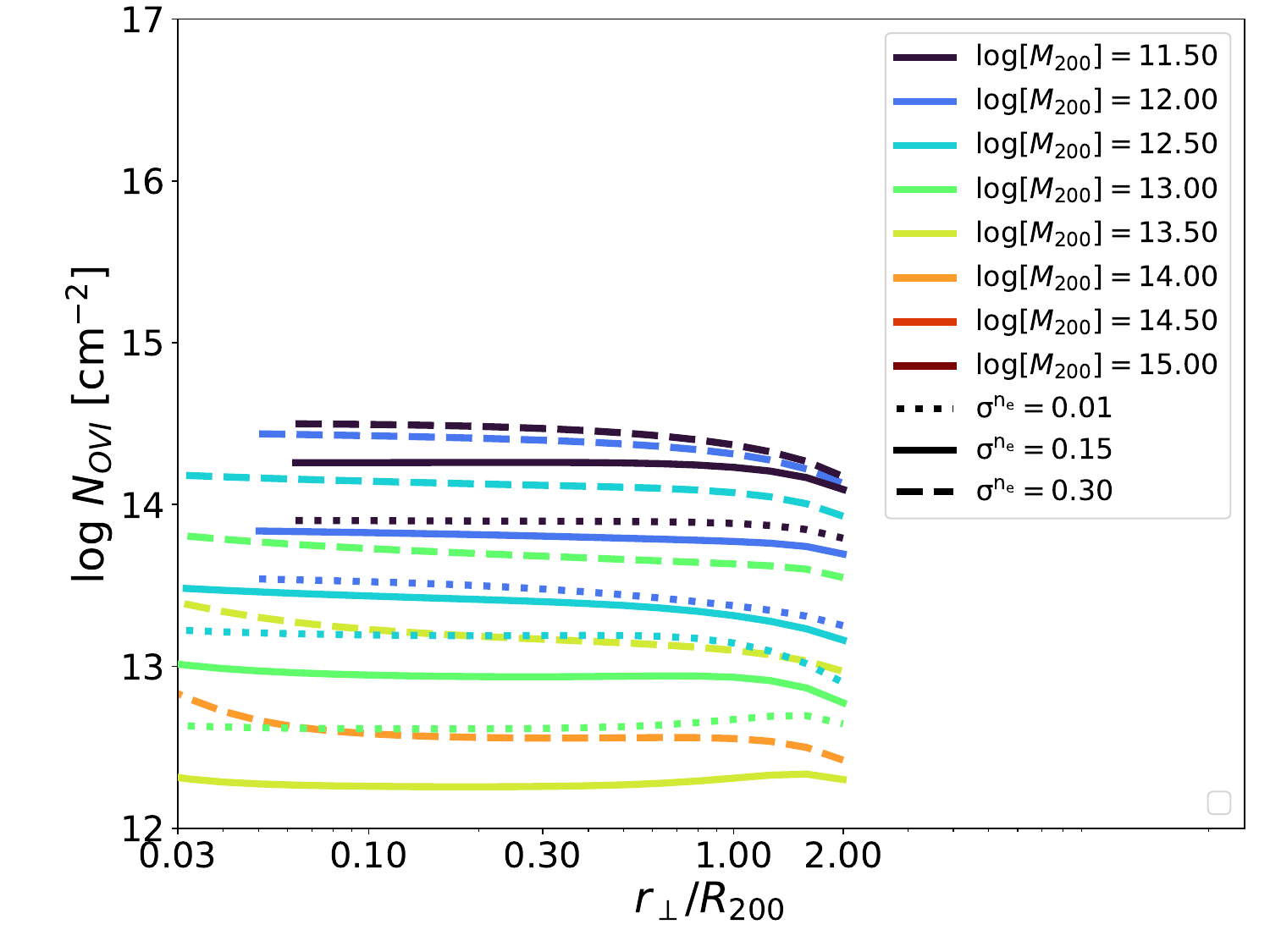}
\caption{Comparisons of $\OVI$ column densities in DPMs and observations. {\it Upper Panel:} $\OVI$ column density from DPMs are shown by lines at different halo masses (distinguished by color) and for models (distinguished by line type) as indicated in the upper right legend. Observations are shown by datapoints as indicated in the lower right legend.  Open symbols are upper limits.  These DPMs have $\sigma^{n_\mathrm{e}}=0.15$. The DPMs are shown at the median redshift of the observational data, $z=0.30$, but the datapoints have a spread in observed redshifts.  $\OVI$ is particularly sensitive to the model and the halo mass. {\it Lower Panel:} $\OVI$ is also sensitive to the density dispersion $\sigma^{n_\mathrm{e}}$ assumed in Model 3 (shown without datapoints for clarity).  This owes primarily to the paired, inverted temperature dispersion given the requirement of constant pressure at a given radius.  Larger dispersion results in less sensitivity to halo mass. }
\label{fig:NOVI}
\end{figure*}

$\OVI$ column densities as a function of fractional impact parameter, $r_\perp/R_{200}$, are plotted in Figure \ref{fig:NOVI}.  Observations compiled by \citet{Tchernyshyov2022} are plotted in the left panel as datapoints according to the lower right legend including COS-Halos \citep{werk2012, Werk2013}, eCGM \citep{Johnson2015}, COS-LRG \citep{Chen2018,Zahedy2019}, COS-GTO-17 \citep{Keeney2017}, COS-GTO-18 \citep{keeney2018}, and CGM$^2$ \citep{Wilde2021}.  Closed symbols are detections and open symbols are upper limits.  We note that we put horizontal error ranges on $r_\perp/R_{200}$ due to the uncertainty in $R_{200}$ derived from $M_{200}$ via SMHM abundance matching (see \S\ref{sec:SMHM}).  The redshift range of $\OVI$ absorbers is $z=0.12-0.60$ with a median of $z = 0.30$, so that DPMs are plotted as lines at $z=0.3$.

Most of the $\OVI$ is formed by collisional ionization (from gas densities that are too high for photo-ionization); therefore it is extremely sensitive to the amount of gas in the $10^{5.5}$ K temperature regime.  The sharp decline in $\OVI$ column densities is reflective of the Virial Temperature Thermometer model of \citet{opp16} and \citet{Wijers2020}, where ion fractions of different oxygen ions peak at CI temperatures that overlap with the virial temperatures of halos (see \S\ref{sec:xrayabs}).  

\paragraph{Observational Comparison and Biases}

In the upper panel of Fig.~\ref{fig:NOVI}, the $M_{200}^{2/3}$ temperature dependence of Model 1 leads to $L^\star$ halos overlapping the $\OVI$ CI band and a faster decline at higher mass.   This is in contrast to Model 2 where $L^\star$ halos barely overlap the CI band.  Model 3 has flatter profiles in the $L^\star$ regime that do not overlap some of the higher observations, although the plotted redshift does not match the datapoints.  

The right panel shows the effect of reducing and increasing $n_\mathrm{e}$ (and inversely $T$) dispersions.  Reducing dispersion to effectively zero ($\sigma^{n_\mathrm{e}}=0.01$) both decreases $N_{\OVI}$ and creates more sensitivity to halo mass, which owes to this model not overlapping the CI band at the lowest temperatures and then rapidly declining at higher mass and hotter temperature.  Adding dispersion increases the $\OVI$ column in $M_{200}=10^{11.5}\;\msolar$ models allowing overlap with the CI band.  Secondly, increasing dispersion reduces $\OVI$ dependence on $M_{200}$ since a larger range of temperatures blur the Virial Temperature Thermometer effect \citep[e.g.,][]{Voit2019_PrecipLimited}.  Nevertheless, the profiles are flat independent of dispersions compared to the observations that rise in the interior.  

We note that there exist many theoretical models of $\OVI$, and some argue that this ion arises from the atmospheric component that arises mainly via $T\sim 10^{5.5}$ K CI $\OVI$ \citep{opp16,Suresh2017,Nelson2018b,Faerman2020}, while others suggest $T\sim 10^4$ K photo-ionized $\OVI$ from the halo gas itself \citep{Stern2018} or extended intergalactic medium \citep[IGM;][]{bromberg2024}, and yet others suggest mixing layer interfaces \citep{Faerman2023} or cooling flows \citep{Heckman2002,Qu2018a}. In addition, star-forming galaxies host more $\OVI$ in the CGM compared to quiescent galaxies with similar stellar masses \citep{Tchernyshyov2023, Qu2024}, suggesting potential enrichment of $\OVI$ gas due to stellar feedback or the role of $\OVI$ gas in quenching galaxies.

\begin{figure*}
\includegraphics[width=0.98\textwidth]{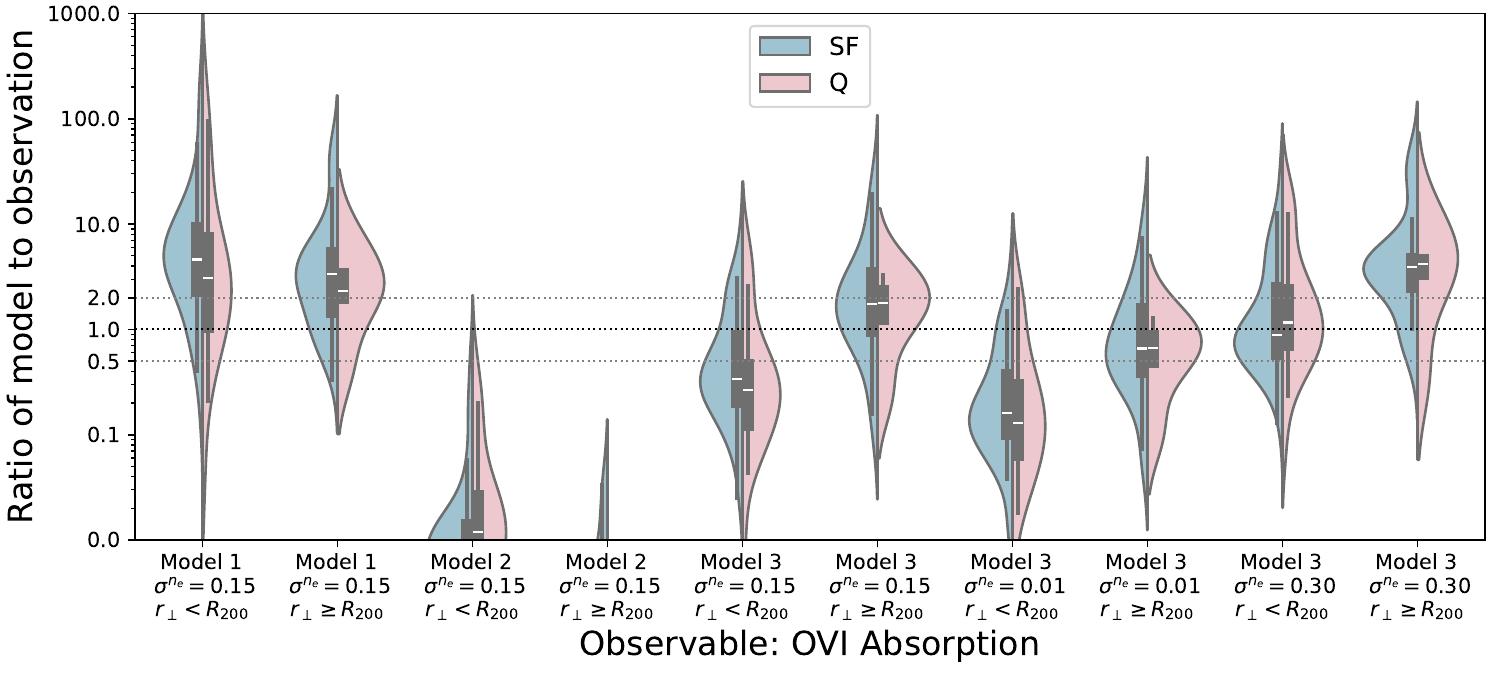}
\caption{Comparison of $\OVI$ column densities of DPMs relative to observed values for five sets of DPM profiles sub-divided into star-forming (SF) and quiescent (Q) galaxies (blue \& pink) and sub-divided again into radii inside and outside $R_{200}$ (two SF/Q sets per model). Model 1-3 at $\sigma^{n_\mathrm{e}}=0.15$ and Model 3 with $\sigma^{n_\mathrm{e}}=0.01$ and $0.30$ are shown from left to right.  These violin plots demonstrate that $\OVI$ is often mis-estimated (e.g. Model 1 (2) often overestimates (underestimates) observations).  Preference is given for models that either match or underestimate observables.  Upper limit observations are treated as detections, leading to artificially lower values on this plot.   Model 3 performs the best inside $R_{200}$, but overestimates $\OVI$ column densities outside $R_{200}$, which may be even more exacerbated by the many upper limits.  Lower $\sigma^{n_\mathrm{e}}$ reduces $N_{\OVI}$. }
\label{fig:violin_OVI}
\end{figure*}

We show DPM to observed column density ratios sub-divided by model, radius, and galaxy type in Figure \ref{fig:violin_OVI}.  We pair observations to the closest halo mass (0.1 dex) and redshift ($\delta z=0.1$) DPM column density profile and match the projected radius.  Model 1 predicts higher columns relative to observations, as expected from Fig.~\ref{fig:NOVI}, while Model 2 shows the opposite.  
The $\OVI$ in Model 3 primarily arises from outside $R_{200}$.  We show three versions of Model 3 with different $\sigma^{n_\mathrm{e}}$.  At $\sigma^{n_\mathrm{e}}=0.15$, Model 3 is intermediate between Models 1 and 2 and has a median ratio of model to observation of 0.25--0.35 inside $R_{200}$.  Model 3 profiles are flat with the greatest deviations at small radii (cf. Fig. \ref{fig:NOVI}, lower panel).  We divide galaxies into star-forming (SF) and quiescent (Q) categories as provided by \citet{Tchernyshyov2022} to demonstrate that there does not exist a significant difference in model fits between the two categories.   

Although we prefer ratios less than one to allow for the presence of a cloud-like component that is not modelled by our DPM, this may be challenging for $\OVI$ as photo-ionized $\OVI$ exists at lower densities that cannot be in thermal pressure equilibrium with the hot phase at small radii. The \citet{opp16} simulations also showed flat profiles from primarily CI gas, which could be remedied with flash ionization from AGN creating residual non-equilibrium photo-ionized $\OVI$ \citep{opp18a}.  

Among Model 3 versions, larger $\sigma^{n_\mathrm{e}}$ create higher $\OVI$ columns for more massive halos.  This trend appears especially discrepant at $r>R_{200}$, where the ratio of model to observation values exceed unity.  For Model 3 to fit these larger radii, a lower dispersion may have to apply.  Alternatively, models with different metallicity relations, e.g., a steeper decline beyond $R_{200}$, may need to be considered.  

We additionally note the uncertainty in using SMHM to estimate both $r_\perp/R_{200}$ and $M_{200}$.  It is more likely that $M_{200}$ and $R_{200}$ are overestimated due to Eddington bias, which could help alleviate some of the tension by moving observed datapoints to larger $r_{\perp}/R_{200}$ and being modelled with lower mass DPM profiles with higher $N_{\OVI}$.  Finally, we explored the FG09 EGB, finding that it leads to slightly higher ($\sim 0.1$ dex) column densities due to more $\OVI$ coming from low densities.

\subsubsection{UV-Derived Pressures} \label{sec:obsPressures} 

\begin{figure*}
\includegraphics[width=0.67\textwidth]{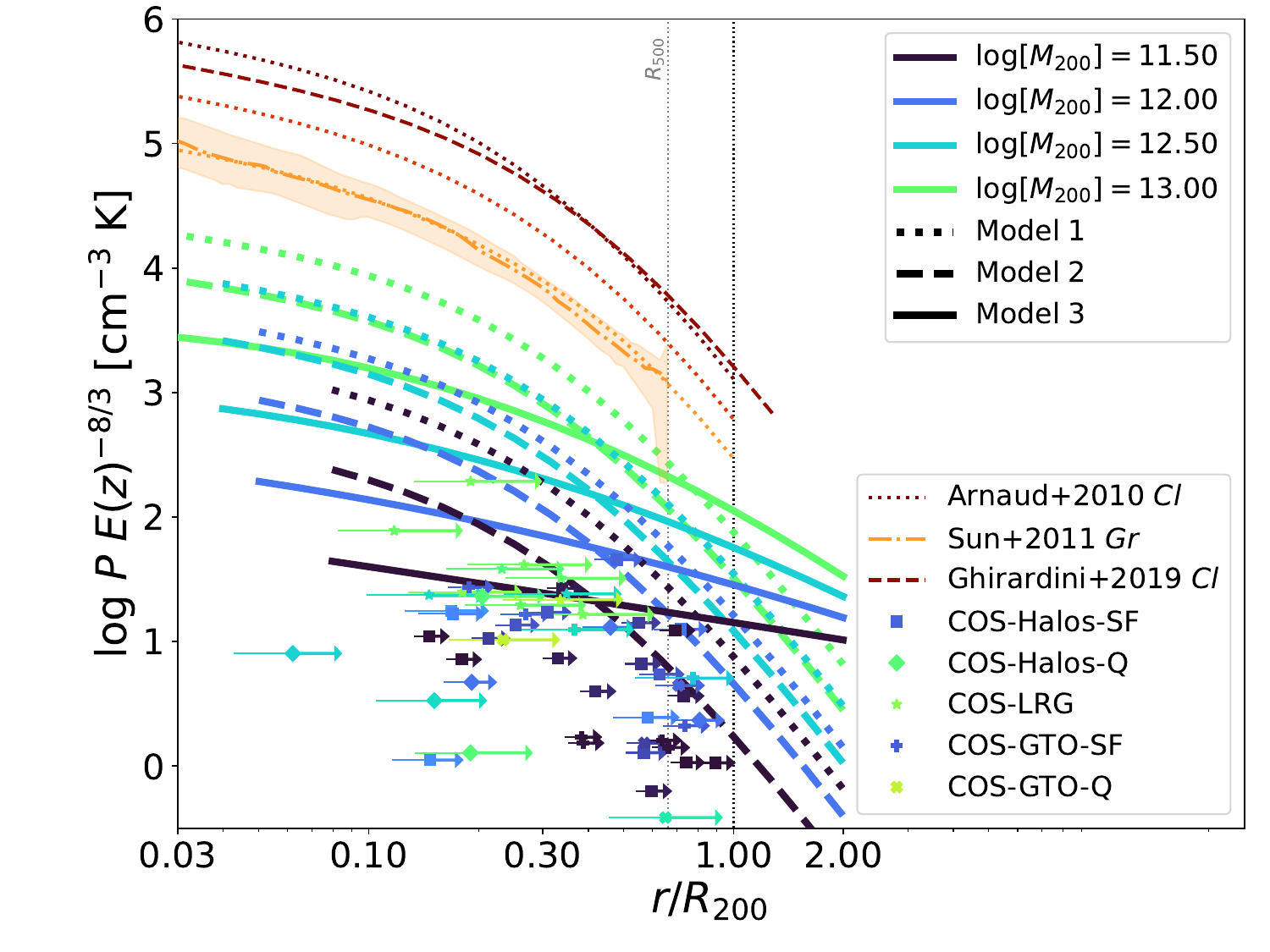}
\caption{Physical electron pressures, including literature data from groups and clusters as lines, which are also plotted in Fig. \ref{fig:pressure_norm}, and UV absorption-derived cloud pressures as datapoints.  $L^\star$ DPM profiles are compared to the observationally-derived UV pressures, with model pressures being significantly higher.  Model profiles are plotted as a function of $r$, although UV pressures are observed for a given $r_{\perp}$ and rightward arrows indicate the possibility of their $r$ values being larger. }
\label{fig:pressure_all}
\end{figure*}

Unnormalized UV-derived pressures are displayed as datapoints in Figure \ref{fig:pressure_all} along with $L^\star$ DPM profiles.  These pressures derived from cool clouds suggest a significant tension with all models and indicate much lower pressures than our atmospheric models predict.  We note that these clouds could reside at larger 3-D radial distance (see \S\ref{sec:mockPressures}), and indicate this possibility by drawing a rightward arrow on these datapoints toward larger $r$.  

\paragraph{Observational Comparison and Biases} 

\begin{figure}
\includegraphics[width=0.49\textwidth]{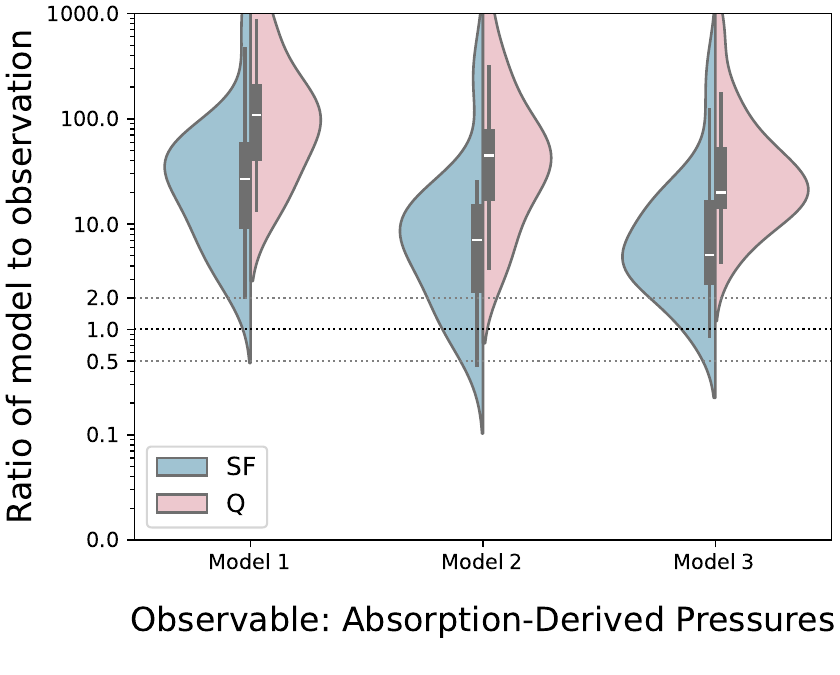}
\caption{Comparison of pressures from DPMs over UV-derived cloud pressures.  Similar to Fig. \ref{fig:violin_OVI}, samples are divided into star-forming (SF) and quiescent (Q), and all pressures are from absorbers with $r_{\perp}<R_{200}$.  UV-derived pressures are significantly lower than all DPM pressure profiles, and the disagreement is greatest for quiescent galaxies.}
\label{fig:violin_Pressures}
\end{figure}

The violin plot in Figure \ref{fig:violin_Pressures} demonstrates that pressures are almost always severely over-estimated by all our models, including Model 3 where the median star-forming (quiescent) galaxy pressure is $5\times$ ($20\times$) too high.  This tension has been noted before by \citet{werk2014}, who found that cool cloud pressures are several orders of magnitude lower than those predicted by the hot, coronal atmospheres of the \citet{Maller2004} model.  That model is best represented by our Model 1, which has its cosmic proportion of baryons mostly in a hot coronal atmosphere and also shows the greatest conflict with observed pressures.  \citet{Faerman2023} constructs a model for cool gas embedded in isentropic warm/hot CGM, where they examine COS-Halos SF galaxies with $\OVI$ detections and find a lower pressure contrast between the phases, typically around 3, with a range of $1-10$.  Their model appears to agree better with our Model 3 that is nearly isentropic around $10^{12}\;\msolar$ halos.  

There are several ways to alleviate this tension.  The first is to use a stronger EGB, which can increase cool cloud densities and pressures and is why we explore the HM05 EGB.  This only partially alleviates the tension as star-forming (quiescent) pressures are $2\times$ ($8\times$) too high (not shown).  

A second fix is to argue that the 3-D radii are much larger than the projected radii, which is the reason for the arrow in Fig. \ref{fig:pressure_all}.  However, we do not favor this explanation since $r\sim r_{\perp}$ is most likely owing to all absorbers having enough ion detections to derive pressures are observed at $r_{\perp}<R_{200}$; therefore it is unlikely that moving them radially outward, often beyond $R_{200}$, would align with the lack of these types of absorbers observed at $r_{\perp}>R_{200}$.  

A third possibility is that the pressure profiles of $L^*$ galaxies drop off near $R_{200}$ more steeply than in Model 3. If their pressure profiles drop below $\sim 10 \, {\rm cm^{-3} \, K}$ at $r \lesssim R_{200}$, then it would be possible for projected clouds with $r \lesssim R_{200}$ to be in pressure equilibrium with the ambient gas.

Finally, non-thermal pressure support in cool clouds is not a mechanism considered in our modelling, but it has been found in simulations with additional pressure support provided by magnetic fields \citep{vandevoort2021}, cosmic rays \citep{Butsky2020, Ji2020,Ruszkowski2023}, and turbulence \citep{Lochhaas2023,sultan2025}.  Using dedicated zoom-in simulations with the EAGLE model, \citet{opp18c} are able to reproduce the observed low metal ion column densities from COS-Halos, but their cool clumps are out of pressure equilibrium due to numerical effects.

\section{Implications from Multi-Band Datasets} \label{sec:implications}

In this section we use the DPM formalism applied to multiple wavebands to discuss physical implications for gaseous halos.  Multiple observables constrain a measurement of pressure, including UV-derived absorber pressures and tSZ stacked profiles, while X-ray emission is also very sensitive.  Likewise, electron density is measured via DMs and kSZ stacks, but all other observables constrain this profile.  We do not discuss metallicity, upon which $\OVI$ and X-ray emission for $L^\star$ halos heavily depend, and which also affects UV-derived absorption pressures.  

\subsection{Pressure Profiles} \label{sec:pressuremismatch}

The much lower absorption-derived pressures around $L^\star$ galaxies (\S\ref{sec:obsPressures}) exhibit the largest tension with the DPMs.  The Compton $y$ parameter profile of \citet{bregman2022,bregman2024} (\S\ref{sec:obstSZ}), which measures thermal pressure, indicates much better agreement.  Furthermore, the eRASS:4 $X_{\rm SB}$ profiles (\S\ref{sec:obseRASS4}) either match or are underestimated by all $L^\star$ models (except in the interior of Model 1, which is ruled out); therefore lower pressures would reduce the X-ray DPM profiles even more.  If we tuned a model to fit absorption-derived pressures, almost every other observable, including $\OVI$ absorption would show a significant mismatch.  Altogether this suggests that these pressures are discrepant, perhaps indicative of non-thermal pressure support in cool clouds. However, we also note that many assumptions, biases, and systematic uncertainties go into measuring absorption-derived pressures, which makes accurate interpretation challenging.  

Finally, we note that quiescent galaxies presumably occupying more massive halos than star-forming galaxies of the same stellar mass \citep{Zu2016} have even more discrepant absorption-derived pressures than from DPM profiles, especially at $r\la 0.3 R_{200}$, which could indicate shallower inner profiles at the transition to the groups regime.  In support of this trend, \citet{Qu2022} finds greater turbulent energy in quiescent versus star-forming halos.  \citet{Zahedy2019} demonstrates that their Luminous Red Galaxy (LRG) cool gas pressures appear to be near equilibrium with the hot gas profiles from \citet{Singh2018}, who derive a lower temperature power-law model for $10^{13}\;\msolar$ halos than Models 2 and 3.  \citet{pratt2021} profiles are also overestimated most in the interior, and the most massive eRASS:4 bin \citep{Zhang2024a} is also over-estimated by DPMs.  Groups of similar mass probed at $z\sim 0.55$ via tSZ \citep{Schaan2021} appear to have underestimated pressures, but mainly at radii outside $R_{200}$.  Our DPMs presented here are continuous functions of halo mass, but simulations find breaks in behavior that can be related to the emergence of black hole feedback \citep{Davies2020,terrazas2020,Voit2024,Medlock2025}.  

\subsection{Electron Density Profiles}\label{sec:densitymismatch}

Compared to pressures, it is more difficult to identify an obvious conflict across observables for density.  For groups, it does appear that the \citet{Schaan2021} kSZ measurement supports a steeper profile like Model 2 at $z\sim 0.55$, while the density-sensitive measures of eRASS:4 groups \citep{Zhang2024a} favor flatter profiles like Model 3 at $z\sim 0.15$, but the different redshifts are not directly conflicting.  In our DPM framework, densities are more ambiguous since we allow density dispersions at a given radii while pressure is fixed.  This creates challenges when diagnosing DPM $\OVI$ predictions for $L^\star$ halos since profile shape and dispersion can create degenerate solutions.  

Future kSZ and DM $L^\star$ constraints will help overcome these degeneracies by returning measurements proportional to the column density of ionized baryons.  These provide an orthogonal constraint to X-ray measurements that have a density-squared dependence; however $L^\star$ halo DMs are currently too unconstraining (\S\ref{sec:obsDM}).  Additionally, both the kSZ Effect and DMs are more readily contaminated by IGM contributions relative to pressure-based measurements; therefore accurate subtraction of the non-halo component is critical.

\section{Future DPM Applications} \label{sec:applications}

We highlight other uses and future directions for the DPM formalism and associated release of observational dataset collections.  The {\tt DPMhalo} GitHub repository at \url{https: /github.com/benopp99/DPMhalo} allows a user to create most of the plots in this paper, apart from the violin plots and Figs.~\ref{fig:tcool} and \ref{fig:SZ_Schaan21}.  We encourage readers to download the {\tt DPMhalo} code and datasets to make their own applications and fits to existing and new datasets.  

\subsection{The {\tt DPMhalo} Module to Model Large Surveys}

We introduce the DPM explicitly to provide a model framework for physical gaseous halo profiles.  The flexibility of the DPM profiles allows a wide range of model profiles for existing and future surveys.  We publish the {\tt DPMhalo} code to offer an integrable modelling module for upcoming large-scale observing campaigns in the X-ray, submillimeter/millimeter, and radio wavebands.   

For the X-ray, the {\it eROSITA} 6-month eRASS:1 dataset is already published \citep{eRASS2024} and the full eRASS:4 dataset will be released in the coming years.  We have mentioned the completed but not yet released X-GAP survey of groups \citep{Eckert2024}, and we anticipate the recently approved {\it XMM} Heritage Program observing the 10~deg$^2$ Euclid Deep Fornax Field (EDFF) for 1000 hours (X-EDFF) that will observe groups and clusters out to $z\approx 2$.  

The deeper sensitivity of the Simons Observatory \citep{simonsobservatory2019} will supersede the ACT kSZ and tSZ results.  Large sky footprint CMB surveys, including the future CMB Stage-4 \citep[CMB-S4;][]{CMBS42016} allow the stacking of subcategories of galaxies observed in spectroscopic surveys probing the CGM and IGrM \citep{Battaglia2017}.  We note the recent release of ACT cross-correlated with DESI LRGs measuring kSZ at an average redshift of $z=0.7$ \citep{Hadzhiyska2024}.  

Lastly, future surveys of localized FRBs from experiments including CHIME \citep{chime2018}, DSA-110 and DSA-2000 \citep{Hallinan_2019}, BURSTT \citep{BURSTT_2022}, and CRAFT \citep{shannon2024} will provide constraining measurements of electron density profiles that distinguish the models presented here.  Similar to kSZ profiles, DMs are heavily affected by the dominant ionized electron reservoir in the IGM.

\subsection{Atmospheric Parameters for CGM \& ICM Studies} \label{sec:atmoparam}

{\tt DPMhalo} models can provide observationally motivated gaseous environments for high-resolution wind tunnel simulations of cool clouds and dwarf galaxies in a Milky Way-like halo \citep[e.g.][]{Tan2023,Zhu2024,abruzzo24}, and infalling galaxies in an ICM setting \citep[e.g.][]{Tonnesen2019,zhu2024a}.  A realistic environment can be created for simulations of dust destruction \citep[e.g.][]{Richie2024} and the survival of Magellanic debris \citep[e.g.][]{Bustard2022}.  Alternatively, simulations of infalling dwarf galaxies \citep[e.g.][]{Zhu2024} and specifically the Magellanic Clouds for the Milky Way \citep[e.g.][]{Salem2015}, can provide constraints on the gaseous halo medium of spiral galaxies via {\tt DPMhalo}.  We avoid using Milky Way-specific constraints in this paper as the data are of a different nature with our location within the Galaxy and the halo gas showing signs of being unique \citep[e.g.,][]{bish21}, but the {\tt DPMhalo} framework is well set up to create realistic generic galactic atmospheres.  

\subsection{X-ray Oxygen Absorption} \label{sec:xrayabs}

\begin{figure*}
\includegraphics[width=0.49\textwidth]{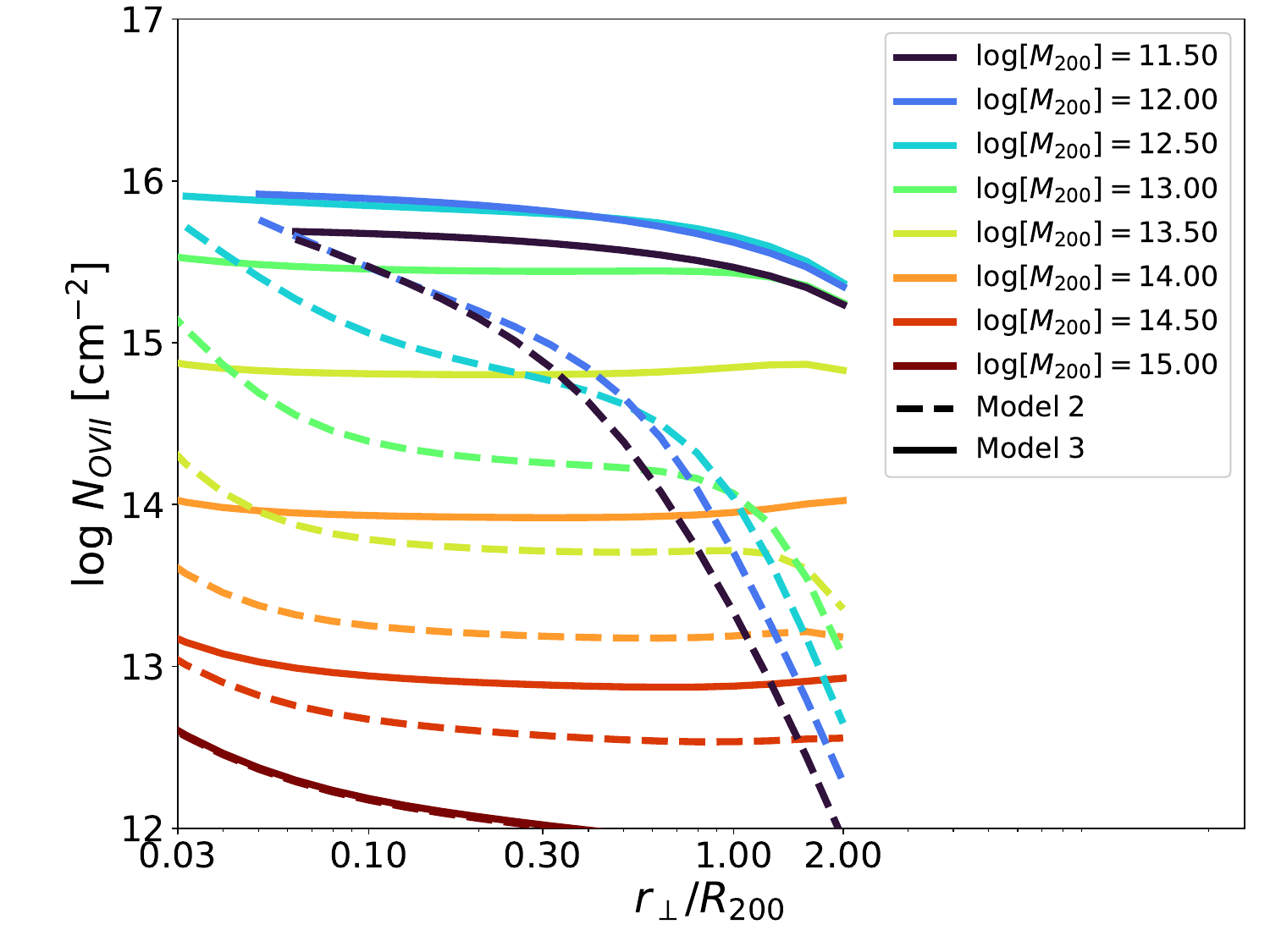}
\includegraphics[width=0.49\textwidth]{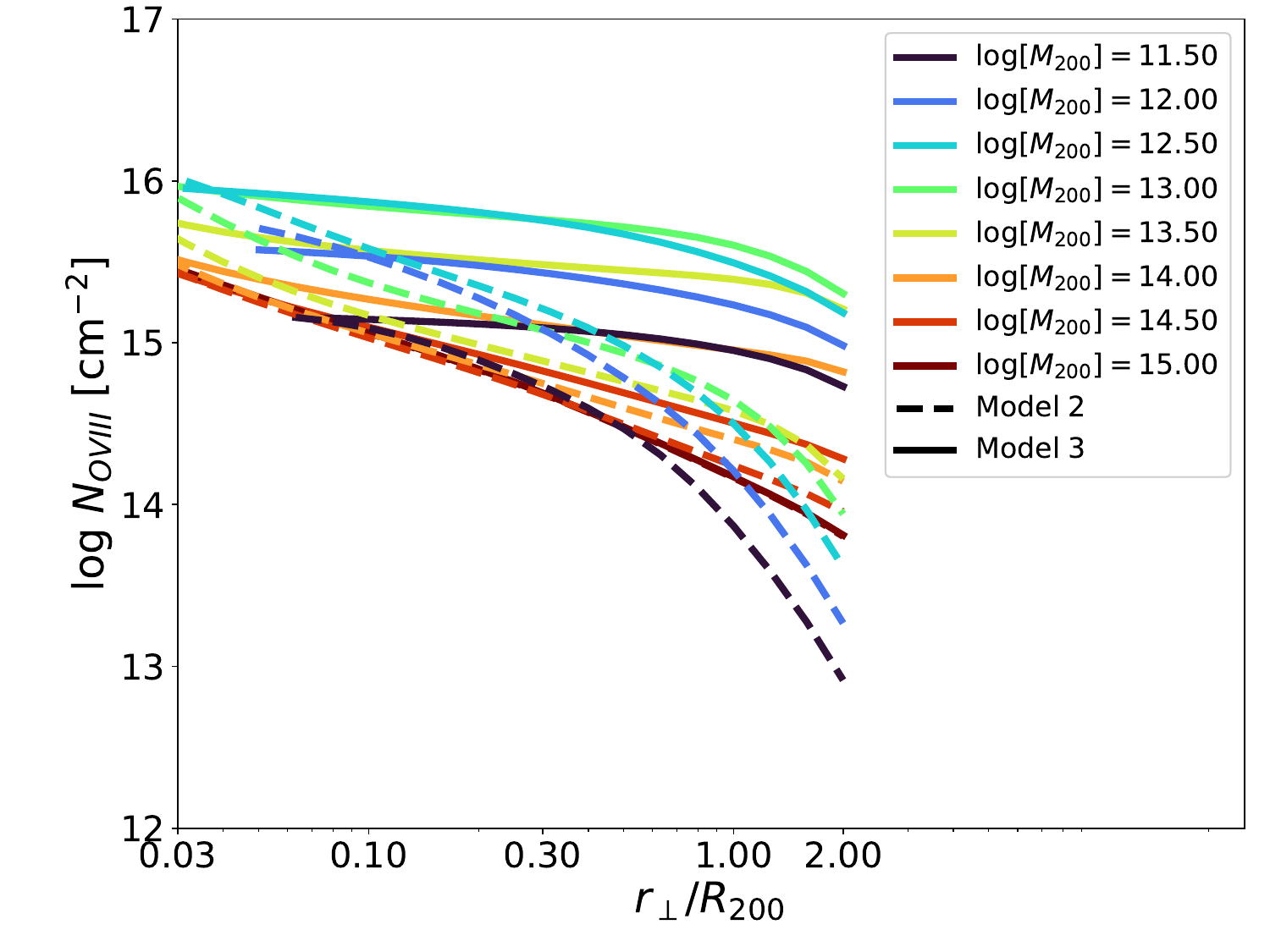}
\caption{$N_{\OVII}$ and $N_{\OVIII}$ radial profiles for Models 2 and 3 at different halo masses for $z=0.3$ assuming $\sigma^{n_\mathrm{e}}=0.15$.  Both models have similar $f_{\rm gas}$ as constrained by existing observations, but their $\OVII$ and $\OVIII$ profiles are strikingly different.  A fundamental purpose of the DPM formalism is to generate predictions like this that can be compared to future observations, in this case a potential X-ray facility that can spectrally resolve oxygen absorption.  }
\label{fig:OVII_OVIII}
\end{figure*}

We output $\OVII$ and $\OVIII$ column density profiles as part of the standard DPM package in anticipation of datasets collected by potential future X-ray absorption line facilities, such as {\it ARCUS} \citep{Smith2023}, as well as absorption observed via emission-line micro-calorimeters including {\it LEM} \citep{bogdan2023} and {\it HUBS} \citep{bregman2023}.  Figure \ref{fig:OVII_OVIII} demonstrates the ability for 
$N_{\OVII}$ and $N_{\OVIII}$ to distinguish halo mass and model profiles.  $\OVII$ in particular has very different predictions for Model 2 and Model 3.  Model 3 shows much stronger absorption at large radii in and around $L^\star$ halos, which should be observable via an ARCUS-like facility \citep{Wijers2020}.  An X-ray spectrometer would be able to constrain both $\OVII$ and $\OVIII$, due to the proximity of their transition energies, in the same objects for a given AGN sight line probing foreground halos.  

\subsection{Dark Matter Profiles}

Along with each model we provide the dark matter halo density in $\gcmc$ using the outputs from {\tt Colossus}.  These profiles are a function of mass and redshift using the chosen mass-concentration relationship \citep{diemer2015}.  They are unaltered by baryonic effects, but a user can apply a model for baryonic contraction (or more accurately expansion for feedback-inflated halos) such as the preservation of angular momentum \citep{blumenthal1986} or another algorithm \citep[e.g.][]{gustafsson2006}.  A user can also calculate baryon to dark matter ratios as a function of radius, as well as test models for hydrostatic equilibrium.

\subsection{Regression Fitting}

We do not perform statistical fits to the observational datapoints, but this should be possible with the release of {\tt DPMhalo} and collected observational datasets.  Our discussion intentionally avoids rigorous statistical regression, focusing more on the nature of observations probing gaseous halos and their potential biases and systematics in \S\ref{sec:obs}. 

A subsequent step would be to apply an MCMC fitting procedure to these datasets, and we make available our collection of observational datapoints, their errors, along with model fits to these datapoints on the {\tt DPMhalo} GitHub repository.  We suggest two possible paths for the treatment of data.  The simplest is to treat observational detections as upper limits for the observations that may arise from the atmospheric plus cloud components, which is relevant for all observations except those that infer pressures (absorption-derived pressures and tSZ).  A more complete approach would be to explicitly add the cool component in combination with DPMs and constrain observations directly.  Examples of these ``cool cloud'' models include {\tt CLOUDFLEX} \citep{hummels2024}, analytic models \citep{Afruni2019,lan2019,Faerman2023,singhbisht2024} and the empirical model from \citet[][see also \citealp{Faerman2025b}]{zheng2024}.  

\subsection{Varying $r_{\rm max}$}

The 3-D radius out to which the physical profiles are generated can significantly alter mock observations, in some cases.  We set the default to $r_{\rm max} = 3 R_{200}$, but allow options spanning $2-5 R_{200}$ in the {\tt DPMhalo} release.  The effects of varying $r_{\rm max}$ are greatest for $\OVI$ absorption and DM profiles, especially for the flatter $L^\star$ profiles of Model 3, while they are nearly negligible for X-ray and tSZ profiles, other than if $r_{\perp}$ approaches $r_{\rm max}$ (not shown).  

\section{Summary}\label{sec:summary}

We introduce the Descriptive Parametric Model (DPM) to generate profiles of gaseous halos and make observational predictions spanning multiple wavelengths.
We have four motivations for this first DPM paper.  First, we develop a mathematical formalism based on the generalized NFW profile that can be applied to spherical gaseous halos spanning clusters to low-mass galaxy CGMs (\S\ref{sec:formalism}).  Our formalism originates from the conventions used for massive halos (clusters and groups) allowing flexible mathematical descriptions of gaseous halos that can be applied both to model large-scale survey science, as well as to provide more robust descriptions for the ambient diffuse halos when modelling the multi-phase CGM.  

Second, we develop the multi-waveband capability to mock observational datasets corresponding to X-ray emission, the Sunyaev-Zel'dovich Effects, electron dispersion measures obtained via FRBs, and UV absorption in a single platform (\S\ref{sec:mock}). 
Up-to-date codes are used to mock X-ray, sub-mm/mm, and UV wavebands.  Systematics, including how the extra-galactic ionizing background can alter our models of UV absorption lines are explored.  We also provide uncertainty estimates related to the calculation of the fractional halo radius using updated treatments for the stellar mass-halo mass relationship.  

Third, we collect observational datasets, introduced in \S\ref{sec:mock} and plotted throughout \S\ref{sec:obs}, and we distribute these datasets along with our code.  Finally, we demonstrate results generated by the DPM code, {\tt DPMhalo}, and compare three idealized models introduced in \S\ref{sec:models} to the observational datasets in \S\ref{sec:obs}.  All our models are calibrated to reproduce $z\sim 0$ galaxy cluster profiles, and they scale differently toward lower halo mass.  Testing how the same set of DPM profiles reproduce multiple observations provides several findings:   

\begin{itemize}
    \item Model 1 (\S\ref{sec:model1}), which has self-similar profiles based on clusters, is the least favorable model given constraints from the X-ray and UV.  Although it has long been known that galactic gaseous halos do not contain hot gas in proportion to clusters, it is informative to compare how this model performs across four wavebands.   

    \item Model 2 (\S\ref{sec:model2}) reduces the gas fraction of hot halos to reproduce the substantially lower gas fractions observed in groups, while preserving the profile shapes from clusters.  $\OVI$ absorption and X-ray emission from galactic halos are substantially reduced to levels far below observations.  While some of these underpredictions are not necessarily problematic and can be compensated for using a separately modelled cool phase, other observational evidence and physical motivations suggest the need to vary radial profile slopes as a function of mass.  

    \item Model 3 (\S\ref{sec:model3}) adopts shallower slopes toward lower halo masses while calibrated to gas fractions like Model 2.  Group X-ray observations support shallower slopes for density, pressure, and entropy, but this model is not constrained directly by these observations.  Model 3 also has longer interior cooling times.  This model often performs the best of the three models, while still being an extrapolation from clusters constrained by several physical criteria at lower mass.
    
    \item No single model can fit all the observed data, and we emphasize overpredictions of observables and mismatched pressures as being most problematic.  We discuss overpredictions including stacked X-ray profiles of massive halos from eRASS:4 (\S\ref{sec:obseRASS4}) and $\OVI$ outside $R_{200}$ (\S\ref{sec:obsOVI}), as well as the underpredicted pressures from absorption lines (tracing cool gas) in tension with SZ-derived pressures for galactic halos (\S\ref{sec:pressuremismatch}).  
        
\end{itemize}

The DPM is {\it descriptive by design} and is presented here as an empirical fitting tool.  The DPM is designed for user-modified applications, which could include calibrating models for well-defined classes of gaseous halos (e.g. quiescent and star-forming halos) or even specific halos (e.g., the Milky Way or M31).  We refer the reader to more potential applications discussed in \S\ref{sec:applications}.  The code is distributed via the {\tt DPMhalo} GitHub at \url{https://github.com/benopp99/DPMhalo}, which also has a link to additional {\tt DPMhalo}-generated plots. \\ 

\section*{Acknowledgements}

The authors thank the anonymous referee for a thorough and fair review of this manuscript that included several important modifications and clarifications.  We also acknowledge Ian McCarthy, Viraj Pandya, Priyanka Singh, Bart Wakker, Fakhri Zahedy, \& John ZuHone for informative discussions that contributed to this manuscript and Jingyao Zhu for doing some initial testing of the code. 

The idea for the DPM project emerged from the KITP program on \textit{Fundamentals of Gaseous Halos}, which was supported in part by the National Science Foundation under Grant No.~NSF PHY-1748958. Further work on it was made possible by the {\it Hubble Space Telescope} Cycle 31 Theory Grant HST-AR-17570.  GMV's contributions were supported in part by grant AAG-2106575 from the NSF. BO, DN, and IM acknowledge support from the NSF grant AST 2206055. JNB gratefully acknowledges support from NASA ADAP grant 80NSSC22K0481 and NSF award 2327438.  YF is supported by the NASA award 19-ATP19-0023 and NSF award AST-2007012.  YZ acknowledges financial support from the European Research Council (ERC) under the European Union’s Horizon 2020 research and innovation program HotMilk (grant agreement No. 865637). YMB acknowledges support from UK Research and Innovation through a Future Leaders Fellowship (grant agreement MR/X035166/1) and financial support from the Swiss National Science Foundation (SNSF) under project 200021\_213076.

\section*{Data Availability}

The {\tt DPMhalo} module is available via Github repository at url{https://github.com/benopp99/DPMhalo}.  Open source packages included in this research include
{\tt Colossus} \citep{diemer2018}, {\tt Mop-c-GT} \citep{Amodeo2021}, {\tt pyXSIM} \citep{Zuhone2016}, {\tt Trident} \citep{Hummels2017}, and {\tt UniverseMachine} \citep{Behroozi2019}.  

\bibliography{ParamModel}{}
\bibliographystyle{mnras}

\end{document}